\documentclass[showpacs,nofootinbib]{revtex4}

\usepackage{hyperref}         %Гипертекстовые ссылки
\usepackage{epsfig,graphicx}
\usepackage{amssymb}
\usepackage{amsfonts}
\usepackage{bm}
\usepackage{graphics}
\usepackage{epsfig}
\usepackage{mciteplus}
\usepackage{epstopdf}

\begin{document}

\title{Hadron interaction with heavy quarkonia}

\author{\firstname{I.~V.}~\surname{Danilkin}}
\email{danilkin@itep.ru} \affiliation{Gesellschaft fur
Schwerionenforschung (GSI) Planck Str. 1, 64291 Darmstadt,
Germany} \affiliation{Institute of Theoretical and Experimental
Physics, Moscow, Russia}

\author{\firstname{V.~D.}~\surname{Orlovsky}}
\email{orlovskii@itep.ru} \affiliation{Institute of Theoretical
and Experimental Physics, Moscow, Russia}

\author{\firstname{Yu.~A.}~\surname{Simonov}}
\email{simonov@itep.ru} \affiliation{Institute of Theoretical and
Experimental Physics, Moscow, Russia}


\newcommand{\be}{\begin{equation}}
\newcommand{\ee}{\end{equation}}

\def\la{\mathrel{\mathpalette\fun <}}
\def\ga{\mathrel{\mathpalette\fun >}}
\def\fun#1#2{\lower3.6pt\vbox{\baselineskip0pt\lineskip.9pt
\ialign{$\mathsurround=0pt#1\hfil ##\hfil$\crcr#2\crcr\sim\crcr}}}
\newcommand{\veX}{\mbox{\boldmath${\rm X}$}}
\newcommand{{\SD}}{\rm SD}
\newcommand{\pp}{\prime\prime}
\newcommand{\veY}{\mbox{\boldmath${\rm Y}$}}
\newcommand{\vex}{\mbox{\boldmath${\rm x}$}}
\newcommand{\vey}{\mbox{\boldmath${\rm y}$}}
\newcommand{\ver}{\mbox{\boldmath${\rm r}$}}
\newcommand{\vesig}{\mbox{\boldmath${\rm \sigma}$}}
\newcommand{\vedelta}{\mbox{\boldmath${\rm \delta}$}}
\newcommand{\veP}{\mbox{\boldmath${\rm P}$}}
\newcommand{\vep}{\mbox{\boldmath${\rm p}$}}
\newcommand{\veq}{\mbox{\boldmath${\rm q}$}}
\newcommand{\veK}{\mbox{\boldmath${\rm K}$}}
\newcommand{\vez}{\mbox{\boldmath${\rm z}$}}
\newcommand{\veS}{\mbox{\boldmath${\rm S}$}}
\newcommand{\veL}{\mbox{\boldmath${\rm L}$}}
\newcommand{\vem}{\mbox{\boldmath${\rm m}$}}
\newcommand{\veQ}{\mbox{\boldmath${\rm Q}$}}
\newcommand{\vel}{\mbox{\boldmath${\rm l}$}}
\newcommand{\veR}{\mbox{\boldmath${\rm R}$}}
\newcommand{\ves}{\mbox{\boldmath${\rm s}$}}
\newcommand{\vek}{\mbox{\boldmath${\rm k}$}}
\newcommand{\ven}{\mbox{\boldmath${\rm n}$}}
\newcommand{\veu}{\mbox{\boldmath${\rm u}$}}
\newcommand{\vev}{\mbox{\boldmath${\rm v}$}}
\newcommand{\veh}{\mbox{\boldmath${\rm h}$}}
\newcommand{\vew}{\mbox{\boldmath${\rm w}$}}
\newcommand{\verho}{\mbox{\boldmath${\rm \rho}$}}
\newcommand{\vexi}{\mbox{\boldmath${\rm \xi}$}}
\newcommand{\veta}{\mbox{\boldmath${\rm \eta}$}}
\newcommand{\veB}{\mbox{\boldmath${\rm B}$}}
\newcommand{\veH}{\mbox{\boldmath${\rm H}$}}
\newcommand{\veE}{\mbox{\boldmath${\rm E}$}}
\newcommand{\veJ}{\mbox{\boldmath${\rm J}$}}
\newcommand{\veal}{\mbox{\boldmath${\rm \alpha}$}}
\newcommand{\vegam}{\mbox{\boldmath${\rm \gamma}$}}
\newcommand{\vepar}{\mbox{\boldmath${\rm \partial}$}}
\newcommand{\vepi}{\mbox{\boldmath${\rm \pi}$}}
\newcommand{{\Mc}}{\mathcal{M}}
\newcommand{\llan}{\langle\langle}
\newcommand{\rran}{\rangle\rangle}
\newcommand{\lan}{\langle}
\newcommand{\ran}{\rangle}

%\documentclass[%
%%prl%
%%,preprint%
% ,secnumarabic%
%%,tightenlines%
%,amssymb, amsmath,nobibnotes, aps, prl ,final]{revtex4}
%\documentclass[aps,12pt,final,notitlepage,oneside,onecolumn,nobibnotes,
%nofootinbib,superscriptaddress,noshowpacs]{revtex4}

\begin{abstract}

Dynamics of hadro-quarkonium system is formulated, based on the channel
coupling of a light hadron $(h)$ and heavy quarkonium $(Q\bar{Q})$ to
intermediate open-flavor heavy-light mesons $(Q\bar{q},\,\bar{Q}q)$. The
resulting effective interaction is defined by overlap integrals of meson
wavefunctions and $(hq\bar q)$ coupling, where $h$ is $\pi, \rho,\omega, \phi$,
without fitting parameters. Equations for hadro-quarkonium amplitudes and
resonance positions are written explicitly, and numerically calculated for the
special case of $\pi \Upsilon(nS)$ $(n=1,2,3)$. It is also shown, that the
recently observed by Belle two peaks $Z_b(10610)$ and $Z_b(10650)$
%at roughly the same energies
%in all $(5,n)$ channels
are in agreement with the proposed theory.
%, and contradict purely molecular explanation.
It is  demonstrated, that theory predicts peaks at the $BB^*, B^*B^*$
thresholds in all available $\pi\Upsilon(nS)$ channels. Analytic nature of
these peaks is investigated, and shown to be due to a  common multichannel
resonance poles close to the $BB^*, B^*B^*$ thresholds. The general mechanism
of these hadro-quarkonium resonances does not assume any molecular or
four-quark (tetraquark) dynamics.

\end{abstract}
\maketitle

\section{Introduction}

It was found in experiment \cite{1,2,3,4} that resonances may
appear in the system of a hadron and heavy quarkonium, which may
be called hadro-quarkonium, see \cite{5} for a review. On
theoretical side the prevalent approaches associate
hadro-quarkonia with molecular or four-quark $(4q)$ states
\cite{p1}-\cite{11}. In the first case hadro-quarkonia are weakly
bound states of two heavy-light mesons of the closest threshold
with interaction tuned to produce loosely bound or virtual states,
and in the $4q$ states thresholds cannot be easily connected  with
$4q$.  However, it will be argued that channel coupling (CC) near
thresholds may play the dominant role in hadro-quarkonium
dynamics, as was shown for heavy quarkonia in our previous papers
\cite{12,13}\footnote{A similar in spirit, but different
technically the so-called rescattering model was developed  and
applied in particular to dipion transitions in bottomonia in
\cite{18', 19', 20', 21', 22'}}. It was shown there, that strong
CC, calculated basically without fitting parameters, shifts the
$2^3P_1$ $(c\bar c)$ pole exactly to the $D\bar D^*$ threshold. In
this way the $X(3872)$ phenomenon was explained using only one
parameter $M_\omega$, which was fixed in previous studies
\cite{13*,13**,13***,17*} and universal input: the string tension
$\sigma$, the current (pole) quark masses, and the strong coupling
$\alpha_s(q)$. Recently $M_\omega $ was found from the first
principles in QCD \cite{22}. It was shown there, that $M_\omega$
can be calculated as the matrix element of the operator   $\sigma
r$, where $\sigma$ is the string  tension and $r$ is the length of
the string. The decay width of $\psi (3770)$ is reproduced in this
way and corresponds to $M_\omega \approx 0.8$ GeV. Our starting
point is the first principle derivation of the CC interaction of
standard heavy quarkonia with open flavor channels, using strong
decay theory \cite{22}. Similarly, in the phenomenon of $X(3872)$,
the systems $\omega\, J/\psi$ and $\rho\, J/\psi$ may take part
with the thresholds near those of $(D\bar D^*+ h.c.)$ states. In
the same way additional pions in the decay vertex appear with the
only extra factor in the denominator $f_\pi = 93$ MeV. As will be
shown below, the strong interaction of pions with $Y(nS)$ mesons
produces charged $Z$-type resonances. Recently in a series of
papers \cite{13*,13**} the CC methods have been successfully
applied to the transitions in systems, containing heavy quarkonia
and pions, or $\eta$ meson and in this way the main features of
experimental pionic spectra in reactions $X' \to X\,\pi\pi$, were
explained, together with kaonic and $\eta$-meson final states
\cite{13***,17*}. Below we extend the formalism of channel
coupling (CC) developed in \cite{12,13}
 to the case of a hadron $h=\pi, \phi,\eta, \rho, \omega,...$
interacting with the $Q\bar Q$ state.

\begin{figure}[t]
  \center{\includegraphics[angle=0,width=0.85\textwidth]{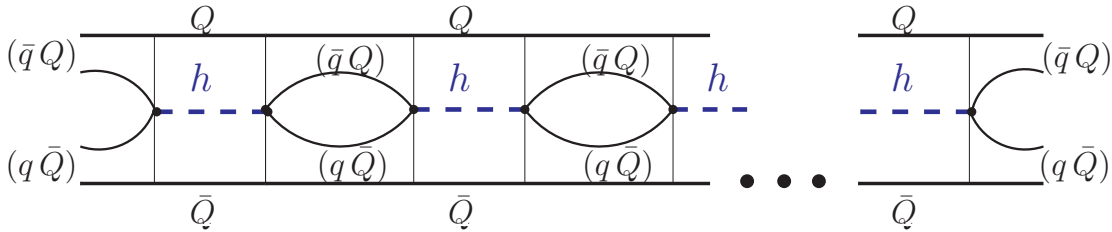}}
  \caption{The chain of transitions of hadro-quarkonia ($h+Q\bar{Q}$) and pair of heavy-light mesons $(Q\bar{q})\,(\bar{Q}q)$. $h$ denotes light hadrons
  $(h=\pi,\phi,...)$.\label{fig:diagram1}}
\end{figure}

We study the interaction and possible poles of hadro-quarkonium amplitudes in
the formalism of \cite{12}. We  assume, that the most important interaction in
hadro-quarkonium is due to intermediate states of heavy-light mesons
$(Q\bar{q})\,(\bar{Q}q)$ (e.g. $DD, DD^*, D^* D^*, D_sD_s,...$ in case of
hadro-charmonium). Therefore one should sum up the whole series of bubbles,
consisting of hadro-quarkonia and heavy-light mesons, as shown in
Fig.\ref{fig:diagram1}. To find poles in amplitudes, one can start and finish
with any state, since poles belong to all channels, i.e. $h\,(Q\bar{Q})$ and
$(Q\bar{q})\,(\bar{Q}q)$. We shall formally study the amplitudes for the
transition of two heavy-light mesons again into the same or other pair of
heavy-light mesons.

It is important to stress, that in our mechanism of
hadro-quarkonium resonances there is no direct interaction neither
in the hadron-quarkonium channel, nor in the channel of two
heavy-light mesons. The only interaction, which generates
resonance  poles, is the CC  interaction, transforming
hadro-quarkonium system into double heavy-light system. Therefore
hadro-quarkonium resonances, predicted in our theory, is a clear
example of CC  resonances, introduced and calculated earlier in
\cite{Bad}. We show below, that direct molecular resonances of
$BB^*$ (if any) are displaced and splitted in hadro-quarkonium in
different hadro-quarkonium channels.

To find the poles, we can  use the so-called Weinberg Eigenvalue Method (WEM),
discussed in detail in \cite{12}. It allows to define  not only poles, but also
resonance wave functions and was successfully applied in \cite{12} to charmonia
states, and in particular to $X(3872)$, in  situation of strongly coupled
channels. It was shown in \cite{12,13} that $X(3872)$ is due to bare
$n=2~^3P_1$ resonance, shifted exactly to the $D_0D^*_0$ threshold by CC and
the detailed experimental form was reproduced in \cite{12,13} with a tiny cusp
at $D_+D_-^*$ threshold and no other bumps. No connection to $\rho\, J/\psi$
and $\omega\, J/\psi$ channels was taken into account in \cite{12,13}, assuming
the corresponding partial widths to be generally small, and  here we  establish
formalism for  these channels, and $\phi\, J/\psi$, which allows to find out,
whether  the CC interaction in these cases is strong enough to produce poles.
The situation with the $\omega\, J/\psi$ channel is especially interesting,
since it contains the resonance of its own, $Y (3940)$ \cite{1}, but in
addition   the decay $X(3872) \to \omega\,J/\psi$,  found  in \cite{14}, (see
\cite{15} for a recent review) suggests that the whole CC system for $X(3872)$
should contain channels $2^3 P_1 (c\bar c)$, $DD^*$, $\omega\, J/\psi$ and
$\rho\, J/\psi$. The CC analysis of this system in another  framework
(Resonance-Spectrum Method) \cite{16}, was done recently in \cite{17}.

A special case of hadro-quarkonium is the pion-quarkonium system, where the
$CC$ interaction vertex is proportional to $1/f_\pi$ and numerically large,
which might support the appearance of $\pi (Q \bar Q)$ resonances. Below we
shall study specifically the case of $\pi\pi$ transitions in the $\Upsilon
(nS)$ states, where these resonances appear in the final states
$\Upsilon(n'S)\, \pi\pi$.

The plan of  the  paper is as follows. In section 2 some basic equations of WEM
in the hadro-quarkonium case are written, and in section 3 those are exploited
to write down exact equation for the possible poles in the general case of
three sectors.  In section 4 the special case of pion-quarkonium system is
treated in detail and $(\pi\Upsilon (nS))$ are found for $n=1,2,3$. In section
5 results of calculations are given, and section 6 contains conclusions,
comparison of molecular and $CC$ dynamics and outlook. Two appendices are
devoted to detailed derivation of decay transition kernel and the form of wave
functions.

\section{Dynamics of strong channel coupling for hadro-quarkonium}

We consider two strongly interacting sectors: sector I with heavy quarkonium
state $(Q\bar Q)$ plus hadron  $h=\pi,\omega, \rho, \eta, \phi$ etc., and
sector II, consisting of two heavy-light mesons $(Q\bar{q})\,(\bar{Q}q)$, in
case of hadrocharmonium, it could be $D\bar D,\, D\bar D^*,\, D^* \bar D^*,\, D
\bar D_1,\, D_s \bar D_s$ etc.

It is important, that we neglect interaction between any white objects,
considering the limit of large $N_c$. It means, that in our treatment there is
no direct interaction between hadron and heavy quarkonium, as well as between
heavy-light mesons. Justification of this approximation can be found in the
fact known from $NN$ interaction, that the main part of long-range forces
between white objects comes from the exchange of one  pion or a pair of
correlated pions, which in case of deutron yields a small binding energy.
However heavy quarks in heavy-light mesons do not contribute in this process
and hence one-pion exchange in the system of two heavy-light mesons should be
much smaller, that in the NN system. This is also supported by the fact, that
$\Lambda N, \Sigma N$ and $\Lambda\Lambda, \Sigma\Sigma$ interactions are
relatively weaker, that the $NN$ interaction. Therefore from our point of view
in the molecular models of exotic charmonia one should take into account that
much stronger attraction near the threshold occurs due to $CC$ interaction
between sectors I and II. One more support of this comes from our recent study
of $X(3872)$ dynamics in \cite{12,13}, where we have shown, that $CC$ alone
strongly shifts $2^3P_1$ $c\bar c$ level  by $ \sim
 60$ MeV to the $D_0\bar D^*_0$ threshold at 3872 MeV.

\begin{figure}[t]
  \center{\includegraphics[angle=0,width=0.3\textwidth]{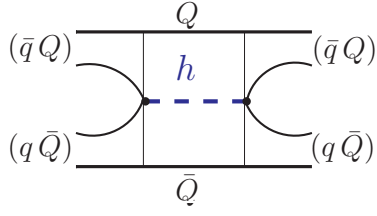}
    \caption{The diagram of the hadron interaction  $V^{(h)}_{n_2n_3; n'_2n'_3}$.\label{fig:diagram2}}}
\end{figure}

As was shown in \cite{12}, to study dynamics of $CC$ in our case, one can
reduce problem to  the one-channel case, where another channel enters via the
$CC$ interaction $V_{aba}, ~{\rm and}~ a,b$ refer to sectors I,II respectively.
If one is interested only in the possibility of bound states or resonances due
to $CC$, one can start with any channel, and we shall work mostly in channel II
and consider the amplitudes shown in Fig.\ref{fig:diagram1} which are generated
by the interaction $V_{212}$. This interaction in the formalism developed in
\cite{12} can be written in momentum space as the amplitude of the loop
diagram, shown in Fig.\ref{fig:diagram2}, with hadron $(h)$ and quarkonium
$(Q\bar Q)$ in the $n$-th state
\begin{equation} V^{(h)}_{n_2 n_3,\,n'_2n'_3} (\vep, \vep', E) = \sum_{n} \int
\frac{d^3 \vek}{(2\pi)^3} \frac{J^{(h)}_{n\,n_2n_3}(\vep, \vek)\,
J^{(h)}_{n\,n'_2n'_3} (\vep', -\vek)}{2\omega_h (\vek)\,(E-E_n(\vek) -
\omega_h(\vek))}\,.\label{1}\end{equation}
with $E_n(\vep) = \sqrt{\vep^2 +M^2_n}$, where $M_n$ is the
position of the n-th bare quarkonium $(Q\bar{Q})$ state. Indices
$n_2 n_3,\, n'_2n'_3$ denote in- and out- quantum states of
heavy-light masons,  the hadron energy is $\omega_h (\vek) =
\sqrt{\vek^2+ m^2_h}$, and the overlap matrix elements
$J_{n\,n_2n_3}^{(h)} (\vep, \vek)$ define the probability
amplitude for the transition of two heavy-light mesons $(Q\bar
q)_{n_2}$, $(\bar Q q)_{n_3}$ with quantum numbers $n_2, n_3$ to
quarkonium $n$-th state $(Q\bar Q)_n$ plus hadron $h$. One can
derive $J_{n\,n_2n_3}^{(h)}$ as a matrix element of a hadron
emission operator between wave functions of quarkonium and two
heavy-light mesons

\begin{eqnarray}
J_{n\,n_2n_3}^{(h)} (\vep,\vek) &=& \frac{1}{\sqrt{N_c}} \int \bar
y^{(h)}_{123}\, \Psi^{(n)}_{Q\bar Q}(\veu -\vev)\, e^{i\vep\ver+ i\vek\vex}\,
\psi_{n_2}(\veu-\vex)\, \psi_{n_3}(\vex-\vev)\,
d^3\vex\, d^3(\veu-\vev) \nonumber\\
& =&  \frac{1}{\sqrt{N_c}} \int\frac{d^3q}{(2\pi)^3}\, \bar y^{(h)}_{123}\,
\Psi^{(n)}_{Q\bar Q}(c\,\vep-\frac{\vek}{2}
+\veq)\,\psi^{(n_2)}_{Q\bar{q}}(\veq)\, \psi^{(n_3)}_{\bar{Q} q}(\veq-\vek).
\label{2}
\end{eqnarray}
where $N_c$ is the number of colours, $\ver =c\,(\veu-\vev),~~ c=
\frac{\omega_Q}{\omega_Q+\omega_q}$. We point that the w.f
$\Psi^{(n)}_{Q\bar{Q}},\,\psi^{(n_2)}_{Q\bar{q}},\,\psi^{(n_3)}_{\bar{Q}q}$
in (\ref{8}) are no longer full w.f. of mesons,  but the radial
part
$R^{(n)}_{Q\bar{Q}},\,R^{(n_2)}_{Q\bar{q}},\,R^{(n_3)}_{\bar{Q}q}$
divided by $\sqrt{4\pi}$, while the angular part of the w.f. is
accounted  for in the factor $\bar{y}_{123}$. This transition
kernel  $\bar y^{(h)}_{123}$ contains a coupling constant $g^q_h$
of hadron with quark pair $(q\bar q)$, entering the hadron
string-breaking $(h q\bar q)$ Lagrangian
\begin{equation} \mathcal{L}_h = \int \bar \psi\, g^q_h\hat e\, \psi\,
\frac{e^{ikx}\,d^4x}{\sqrt{2\,\omega V_3}},\label{3}\end{equation}
and another part, which comes from the Dirac trace of $\gamma$ matrices
corresponding to the vertices in state  $(Q\bar Q)_n$ and $(Q\bar q)_{n_2},\,
(\bar Q q)_{n_3}$.

To obtain the full vertex $\bar y^{(h)}_{123}$  in (2), one can use either the
$(4\times 4)$ form given in \cite{13*,13**} or else the $(2\times 2)$ form for
wave functions and vertices, introduced in \cite{12}, Appendix B. Exact
expressions for $\bar y^{(h)}_{123}$ are  given in Appendix 1 for the
convenience of the reader. In this way for the $n_1$ state of quarkonia and
vector hadron one can write similarly to (B.3)

\begin{equation} \bar y^{(h)}_{123} =tr \{\Gamma^{(n_1)}_{red}\,
\Gamma_{red}^{(n_2)}\, (\mathbf{e} \vesig)\, g_h\,
\Gamma_{red}^{(n_3)}\}\label{4}\end{equation} and \begin{equation}
\Gamma_{red}^{(n)} (D) =\frac{1}{\sqrt{2}},~~ \Gamma_{red}^{(n)}(D^*)
=\frac{\sigma_k}{\sqrt{2}}, ~~\Gamma_{red}^{(n)}(1^{--}(Q\bar Q))
=\frac{\sigma_i}{\sqrt{2}}.\label{5}\end{equation}
Therefore in the case $1^{--}(Q\bar Q)+ h\to D\bar D^*,$ one obtains
\begin{equation} \bar y^{(h)}_{123} =i\,g_h\,e_j\,\epsilon_{ijk}
\label{6}
\end{equation}
where $\mathbf{e}$ is the polarization vector of a hadron and $\epsilon_{ijk}$
is the Levi-Civita symbol.

For Nambu-Goldstone bosons $(\pi,\,K,\,\eta)$ the transition kernel was
obtained in a different way in \cite{13*}. Indeed, pions  accompany string
breaking, yielding a coefficient $\frac{M_{\omega}}{f_\pi}\,\gamma_5$ instead
of $g^q_h\hat e$ in (\ref{3}), where $M_\omega$ is calculated via string
tension $\sigma$ in \cite{22}, $M_\omega= 0.8$ GeV. This  is used in section 4
below, details are given in appendices 1 and 2.

To define possible resonance position and wave function it is convenient to use
the Weinberg Eigenvalue Method,  which was extended to the case of coupled
channel problem in \cite{12}. The corresponding equation for eigenfunction
$\Psi_\nu (\ver, E)$ and eigenvalue $\eta_\nu (E)$ can be written as

\begin{equation} H_0\, \Psi_\nu(\ver, E) + \int \frac{V_{212}
(\ver, \ver', E)}{\eta_\nu(E)}\, \Psi_\nu(\ver', E)\, d^3\ver' =E\,\Psi_\nu
(\ver, E)\label{7}\end{equation} with boundary condition $\Psi_\nu (r\to
\infty, E) \sim  \exp(ikr)/r,\, \Psi_\nu (0,E)=const$ and index $\nu$ labels
the discrete eigenvalues and eigenvectors. In the momentum space one can write

\begin{eqnarray} \psi_{n_2n_3} (\vep,E) &=& -\frac{1}{\eta
_\nu(E)}\sum_{n_2'n_3'} \int\frac{d^3\vep'}{(2\pi)^3}\,
G^{(0)}_{n_2n_3}(\vep)\, V^{(h)}_{n_2n_3, n'_2n'_3}(\vep, \vep',
E)\,\psi_{n'_2n'_3}(\vep')\nonumber\\
G^{(0)}_{n_2n_3} (\vep, E) &=& \frac{1}{E_{n_2}(\vep)+E_{n_3}(\vep)-E}\,.
\label{8}
\end{eqnarray}
At this  point one realizes, that Eq.(\ref{8}) can be seriously simplified,
using the structure of the overlap integrals in (\ref{2}). Indeed, it was shown
in \cite{13*,13**} that wave functions of heavy-light mesons $D, D^*$ $(B,B^*)$
can be represented by the Gaussian functions with accuracy of the order of few
percent. In this case the integral in (\ref{2}) factorizes (see Eq.(21) in
\cite{13**})
\begin{equation}
J_{nn_2n_3}^{(h)} (\vep,\vek) \equiv
\frac{1}{\sqrt{N_c}}\,\varphi^{(h)}_{n_2n_3}(\vek)\,
\chi^{(h)}_{nn_2n_3}(\vep)\,. \label{9a}
\end{equation}
Moreover, it appears, that $\varphi^{(h)}_{n_2n_3}(\vek)$ are
almost identical for the first two states of heavy-light mesons
(e.g. $B,\, B^*$) and are very close for the next two states (e.g.
$B_s,\, B_s^*$), hence we put by simplify $\varphi^{(h)}_{n_2n_3}$
to $\varphi^{(h)}$.

 As one can see in (\ref{1}), the integral on
the r.h.s. also factorizes, when (\ref{9a}) is used, and one can write

\begin{equation}
V^{(h)}_{n_2n_3, n'_2n'_3} (\vep, \vep', E) =-
\sum_n\,\frac{1}{N_c}\,\chi^{(h)}_{nn_2n_3}(\vep)\,
\chi^{(h)}_{nn'_2n'_3}(\vep')\,K_n(E)\,,\label{10a}
\end{equation}
where notation is used
\begin{equation}
K_n(E)= \int \frac{d^3\vek}{(2\pi)^3} \frac{\varphi^{(h)}(\vek)\,
\varphi^{(h)}(-\vek)}{2\omega_h(\vek)\,(E_n(\vek) + \omega_h(\vek)
-E)}\,.\label{11a}
\end{equation}
Insertion of (\ref{10a}) in (\ref{8}) immediately yields
\begin{equation}
\psi_{n_2n_3} (\vep, E) = \frac{1}{N_c}\,\frac{G^{(0)}_{n_2n_3}(\vep,
E)}{\eta_\nu (E)} \sum_{n} K_n(E)\, \chi^{(h)}_{nn_2n_3} (\vep)\sum_{n'_2n'_3}
\int\frac{d^3\vep'}{(2\pi)^3}\, \chi^{(h)}_{nn'_2n'_3}(\vep')\,
\psi_{n'_2n'_3}(\vep',E)\label{12a}\end{equation} which finally leads to a
system of algebraic equations
\begin{eqnarray}
\Lambda_{n'} (E) &=& \frac{1}{\eta_\nu(E)} \sum_n \zeta_{n'n}(E)\,
K_n(E)\, \Lambda_n(E),\label{13a} \\
\Lambda_n(E) &\equiv& \sum_{n_2n_3} \int \frac{d^3\vep}{(2\pi)^3}\,
\chi_{n\,n_2n_3}^{(h)}(\vep)\, \psi_{n_2n_3}(\vep, E)\,, \nonumber
\end{eqnarray}
where we have defined in (\ref{13a})
\begin{equation}
\zeta_{n\,n'} (E) = \frac{1}{N_c} \sum_{n_2n_3}\int\frac{d^3\vep}{(2\pi)^3}\,
G_{n_2n_3}^{(0)}(\vep,E)\,\chi_{n\,n_2n_3}^{(h)}(\vep)\,\chi^{(h)}_{n'n_2n_3}(\vep)\,.\label{14a}
\end{equation}
So we obtain
\begin{equation}
\det \Big[\eta_\nu (E)\,\delta_{nn'} -
K_{n}(E)\,\zeta_{nn'}(E)\Big]=0\label{15a}
\end{equation}
and one can look for poles $E_p$, setting $\eta_\nu(E=E_p)=1$.

%%%%%%%%%%%%%%%%%%%%%%%%%%%%
%%%%%%%%%%%%%%%%%%%%%%%%%%%%

Equivalently one can define  amplitude in the sector I, $h+(Q\bar Q)$, in which
case the interaction ``potential'' $V_{121}$ assumes the form
\begin{equation} V_{nn'}^{(h)} (\vek, \vek', E) = \sum_{n_2 n_3} \int
\frac{d^3\vep}{(2\pi)^3} \frac{J^{(h)}_{nn_2n_3}(\vep, \vek)\,
J^{(h)}_{n'n_2n_3}(\vep,\vek')}{E-H_0^{(n_2n_3)} (\vep)}\label{9}\end{equation}
and the WEM equation in sector I is
\begin{equation}
\psi^{(\nu)}_n (\vek, E) =- \int \frac{d^3\vek'}{(2\pi)^3\,2\omega
(\vek')}\,\, G^{(0)}_n(\vek, E)\, \frac{V_{nn'}^{(h)}(\vek, \vek',
E)}{\eta_\nu(E)}\, \psi^{(\nu)}_{n'}(\vek',E).\label{10}
\end{equation}
One can see from (\ref{9}), that $V_{121}^{(h)} $ is attractive and real for
$E$ below the $D\bar D\,(D\bar D^*)$ threshold; and $G_n^{(0)}$ is
\begin{equation}
G_n^{(0)}(\vek, E)=\frac{1}{H_0(\vek)-E}=
\frac{1}{E_{n}(\vek)+\omega_h(\vek)-E}. \label{11}
\end{equation}
The total Green's function in sector I has the form (see \cite{12} for
discussion and details)
\begin{equation} G^{(I)} (1,2;E) = \sum_\nu\frac{\psi^{(\nu)}_n(1,E)\,
\psi^{+(\nu)}_n(2,E)}{1-\eta_\nu (E)}\label{12}\end{equation}
and near the resonance $\eta_\nu(E)$ has the form \begin{equation} \eta_\nu(E)
= 1+\eta' (E_0 -\frac{i\Gamma}{2})\,\left(E-E_0
+\frac{i\Gamma}{2}\right)+...\label{13}\end{equation} Finally, one can define
the $t$-matrix
\begin{equation} t=\hat V-\hat V\, G\,\hat V\label{14}\end{equation}
which in the WEM can be written as \begin{equation} t(\vek, \vek',E) =-
\sum_{\nu} \frac{\eta_\nu(E)\, a_\nu(\vek, E)\,
a_\nu(\vek',E)}{1-\eta_\nu(E)}\label{15}\end{equation}
where
\begin{equation} a_\nu(\vek, E) =(H_0(\vek)
-E)\,\Psi^{(\nu)}_n(\vek,E)\label{16}\end{equation}

\section{A general case of three coupled sectors}

Till now only the connection of given channel $h+ (Q\bar Q)_n $ to
$(Q\bar{q})\,(\bar{Q}q)$ was considered. A  more interesting situation can
occur, when one adds also excited channels of $(Q\bar Q)_{n'}$. An example of
the physical situation of this kind is given by the $2^3P_1(c\bar c)$ state
connected to $\omega\, J/\psi$ via the $DD^*$ channel. Hence in this case one
has to consider three sectors: as before sectors  I and II refer to $h\,(Q\bar
Q)_{n}$ and $(Q\bar{q})\,(\bar{Q}q)$ respectively and sector III refers to
$(Q\bar Q)_{n'}$ states. One again writes equation for the wave function in
sector II as in Eq. (\ref{8}), but the interaction $V_{n_2n_3,n'_2n'_3}$ now
consists of two terms:
\begin{equation}
V_{n_2n_3,n'_2n'_3}(\vep,\vep', E)= V_{n_2n_3,n'_2n'_3}^{(h)} (\vep,\vep',
E)+V_{n_2n_3,n'_2n'_3}^{(Q\bar Q)} (\vep,\vep', E) \label{15*}
\end{equation}
where we add to the potential $V^{(h)}$ in Eq.(\ref{1}) another kernel, $
V^{(Q\bar Q)}$ of the following form (cf. Eq.(26) of \cite{12}).
\begin{equation}
V_{n_2n_3,n'_2n'_3}^{(Q\bar Q)} (\vep,\vep', E)=\sum_n\frac{
J^+_{nn_2n_3} (\vep) J_{nn'_2n'_3}(\vep')}{E-M_n}. \label{16*}
\end{equation}
Here $J(\vep)$ is obtained from $J^{(h)}(\vep, \vek)$ putting $\vek=0$ and
replacing $\bar y^{(h)}_{123}$ by $M_\omega\,\bar y_{123}$, where $\bar
y_{123}$ is  given in \cite{12} for different states, and $M_\omega$ is a fixed
parameter for all charmonia and bottomonia states used in
\cite{12}-\cite{13***}, $M_\omega =0.8$ GeV, in actual calculations one reduces
the fully relativistic vertex $M_\omega\,\bar y_{123}$ to the two-component
spinor form, convenient for nonrelativistic form of participating
wave-functions $M_\omega\, \bar y_{123}=\gamma\, y^{red}_{123}$, and $\gamma
=1.4$, see appendix C of \cite{12} for details.

The resulting equation for $\eta_\nu(E) $ has the same form as in (\ref{13a}),
but now $\Lambda_n$ and $K_n$ are columns and $\zeta$ is a matrix in $n$ and
$n'$ indices. Fixing $n'$ and denoting  a single channel $n_2 n_3 =
n'_2n'_3\equiv 1,$ one arrives at the equation
\begin{equation} \det\left(
\begin{array}{ll}
1-\frac{1}{\eta_\nu}\, \zeta_{11}\, K_1& - \frac{\zeta_{12}}{\eta_\nu\,(M_{n'}-E)}\\
-\frac{1}{\eta_\nu}\,\zeta_{21}\,K_1& 1-
\frac{\zeta_{22}}{\eta_\nu(M_{n'}-E)}\end{array}\right)=0,\label{17*}\end{equation}
where
\begin{eqnarray}
\zeta_{12}&=&\zeta_{21}\equiv \int \frac{d^3p}{(2\pi)^3}\,\,
G^{(0)}_{n_2n_3}(\vep, E)\, \chi_{nn_2n_3}(\vep)\,
J_{n'n_2n_3}(\vep)\nonumber\\
\zeta_{22} &\equiv& \sum_{n_2n_3}\int
\frac{d^3\vep}{(2\pi)^3}\,\,G^{(0)}_{n_2n_3}(\vep, E)\,
J^2_{n'n_2n_3}(\vep)\,.\label{18*}\end{eqnarray}
Solution of Eq.(\ref{17*}) for $\eta_\nu (E=E_p)=1$ gives the pole positions
$E_p$. In particular, one can derive how the original $(Q\bar Q)$ pole is
shifted due to two effects: 1) $CC$ to the sector II of two heavy-light states
$(Q\bar{q})\,(\bar{Q}q)$ 2) $CC$ due to the hadroquarkonium states -- sector I.

It is clear in (\ref{17*}), that the resulting $\eta_\nu(E)$ contains threshold
singularities from sectors II and I and the pole at $M_n$ in the limit of small
$CC$. In the spirit of calculations in \cite{13} and using (\ref{15}) one can
define the probability of transition from sector I to sector II as the
absorptive part of $t$ (\ref{15}) due to sector II. It yields
\begin{equation} \sigma_{I~II} \sim \frac{\frac{1}{2i}\,
\Delta_{II} \eta_\nu(E)}{|1-\eta_\nu(E)|^2}\label{28}\end{equation}

The total width $\Gamma$ of the resonance, originating from the state $(Q\bar
Q)_{n'}$ in sector III is obtained  from the expansion (\ref{13}) for small
$\Gamma$, and from the position of the pole $E_p$ in the equation
$\eta_\nu(E_p) =1$, where $\eta_\nu$ is found from (\ref{17*}), for arbitrary
$\gamma$. The partial widths of the resonance, originating from sector III,
corresponding to channels in sectors I and II are  proportional to absorptive
parts of $\eta_\nu(E)$ on the cuts, starting from thresholds of sectors I and
II.  The concrete examples of $X(3872)$ connected to $\omega J/\psi$ and $\rho
J/\psi$ states will be given elsewhere.

\section{ Pion-quarkonium resonances}

As a specific example of the general formalism  in section 2, we consider here
the pion interaction with heavy quarkonium.  To simplify matter we shall not
use WEM in this section, writing all expressions in standard form, since we
shall not use the notion of resonance wave function for pion quarkonium. The
corresponding interaction term $V^{(\pi)}_{n_2n_3, n'_2n'_3} (\vep, \vep',E)$
is given  in (1), while $J^{(\pi)}_{nn_2n_3}$ is defined in (2). However now,
in contrast to the vector hadron case of (\ref{6}), one has instead for pion
emission the same vertex, which was derived in \cite{12,13*,13**}. For $DD^*$
or $BB^*$ one has
\begin{equation}
\bar y^{(\pi)}_{123^*}
=\frac{M_\omega}{f_\pi}\frac{i\,\delta_{ik}}{\sqrt{2}}\,, \label{29}
\end{equation}
while for $D^*D^*(B^*B^*)$ one obtains
\begin{equation}
\bar y^{(\pi)}_{12^*3^*}=-\frac{M_\omega}{f_\pi}\frac{e_{ikl}}{\sqrt{2}}\,.
\label{30}
\end{equation}
Note, that indices $i,k,l$ refer to the polarization states of initial and two
final vector mesons. Also, the shorthand notation $DD^*$ implies
$\frac{1}{\sqrt{2}}\,(D\bar D^*\pm D^*\bar D)_I$ for isospin  $I$state.
Finally, as in (\ref{3}), (\ref{6}) the factor $\frac{1}{\sqrt{2\omega\, V_3}}$
is taken into account in the pion phase space integral over
$\frac{d^3\vek}{(2\pi)^3}$.

With these $s$-wave-type kernels (\ref{29}), (\ref{30}) and SHO wavefunctions
in (\ref{2}) one can write the factorized form for $J^{(\pi)}_{nn_2n_3}$
\begin{equation}
J^{(\pi)}_{nn_2n_3}(\vep,\vek) =
\frac{1}{\sqrt{N_c}}\,\varphi^{(\pi)}_{n_2n_3}(\vek)\,\chi_{nn_2n_3}(\vep)\,,
\label{31}
\end{equation}
where
\begin{equation} \varphi^{(\pi)}_{n_2n_3}(\vek) =
e^{-\frac{\vek^2}{4\beta^2_2}},~~ \chi_{nn_2n_3}(\vep) =\bar
y^{(\pi)}_{123}\,e^{-\frac{(c\vep)^2}{\Delta_n}}\,I_{nn_2n_3}(c\vep)\label{32}\end{equation}
and $I_{nn_2n_3}(\vep)$ for SHO functions for heavy-light mesons with
$n_2=n_3=1$ (which gives 95\% accuracy for $B,B^*$ and $D,D^*$ mesons
\cite{12,13*,13**}) is
\begin{equation} I_{n_{11}} (\vep)= 2\,\tilde c_n\, (-)^{n-1}\,
\frac{(2n-1)!}{(n-1)!}\,\,\Phi(-(n-1),\,\frac32,\,\mathbf{f}^2)\,
\frac{y^{n-1}}{(2\sqrt{\pi})^3}
\left(\frac{2\beta^2_1\beta^2_2}{\Delta_n}\right)^{3/2}\label{33}\end{equation}
where all constants are defined via SHO parameters of wavefunctions,
participating in the overlap integral (\ref{2}), see  Appendix 2.

We note, that $\varphi^{(\pi)}_{n_2n_3}(\vek) \equiv \varphi (\vek)$ with a
good accuracy does not depend on  $n_2n_3$,  when $n_2,\, n_3$ run over  a pair
of  indices $B, B^*$ (or $D,\,D^*$ etc.), since the corresponding wavefunctions
are very similar. Therefore the kernel $K_n(E)$ in (\ref{11})   also does not
depend on indices  $n_2n_3$ and can be written as
\begin{equation}
K_n(E) =\int \frac{d^3\vek}{(2\pi)^3}
\frac{e^{-\frac{\vek^2}{2\beta^2_2}}}{2\omega_\pi(\vek)\,(E_n (\vek)+
\omega_\pi(\vek) - E)}\,. \label{34}
\end{equation}
%
%and $E_n (\vek) = \sqrt{ M^2_n + \vek^2}, \omega_\pi (\vek) =
%\sqrt{m^2_\pi + \vek^2}$.
%
Defining as in (\ref{14a})
\begin{equation}
\zeta_{nn'} (E) = \frac{1}{N_c}\sum_{n_2n_3} \int \frac{d^3\vep}{(2\pi)^3}
\frac{\chi_{nn_2n_3}(\vep)\, \chi_{n'n_2n_3}(\vep)}{E_{n_2}(\vep) + E_{n_3}
(\vep) - E}\,, \label{35}
\end{equation}
one has a system of equations (\ref{10}), from which one defines resonance
energy  \cite{12}
\begin{equation}
\det \Big[1 - {\hat K (E)}\,\hat\zeta (E)\Big] =0, \label{36}
\end{equation}
where $(\hat K)_{nn'} = K_n \delta_{nn'}$ and $(\hat \zeta)_{nn'} =
\zeta_{nn'}$.

Note, that $\chi_{nn_2n_3}$ depends on polarization states  of all particles,
and that of $n$ (index $i$ in (\ref{29}), (\ref{30})) can be fixed, while one
should sum up in (\ref{35}) over spin and isospin projection of particles $n_2,
n_3$.

At the end of this section we consider the contribution of pion-quarkonium
resonance into production crossections of the final state $(Q\bar
Q)_{n'}\,\pi\pi$. In the zeroth approximation the amplitude for the transition
$(Q\bar Q)_n \to (Q\bar Q)_{n'}\, \pi\pi$ was calculated  in \cite{13*, 13**,
13***}.

$$ w_{nm}^{(\pi\pi)} (E) = \sum_k
\frac{d^3p}{(2\pi)^3}\frac{J_{nn_2n_3}^{(1)}(\vep,\vek_1)\,J^{*(1)}_{mn_2n_3}
(\vep,\vek_2)}{E-E_{n_2n_3}(\vep)-E_\pi(\vek_1)}+ (1\leftrightarrow 2)$$

$$-\sum_{n'_2n'_3}
\frac{d^3p}{(2\pi)^3}\frac{J_{nn'_2n'_3}^{(2)}(\vep,\vek_1,\vek_2)\,J_{mn'_2n'_3}^*
(\vep)}{E-E_{n'_2n'_3}(\vep)-E(\vek_1,\vek_2)}-$$
\begin{equation}-\sum_{k^{\prime\prime}}
\frac{d^3p}{(2\pi)^3}\frac{J_{nn^{\prime\prime}_2n^{\prime\prime}_3}(\vep)\,J^{*(2)}_{mn^{\prime\prime}_2n^{\prime\prime}_3}
(\vep,\vek_1,\vek_2)}{E-E_{n^{\prime\prime}_2n^{\prime\prime}_3}(\vep)}\label{37}\end{equation}
Here $J_{nn_2n_3}^{(1)}\equiv J^{(\pi)}_{nn_2n_3}(\vep, \vek)$ and
\begin{equation}
J_{nn_2n_3}^{(2)}(\vep, \vek_1, \vek_2) \equiv\frac{1}{\sqrt{N_c}} \,\bar
y^{(\pi\pi)}_{123}\, e^{-\frac{\big(\vek_1+\vek_2\big)^2}{4\beta^2_2}}\,
e^{-\frac{(c\,\vep)^2}{\Delta_n}}\, I_{nn_2n_3}(c\,\vep);\label{38}
\end{equation}
with
\begin{equation} \bar y^{(\pi\pi)}_{123} =\frac{M_\omega
\vepi_1\vepi_2}{f^2_\pi} \frac{\bar y_{123}}{\sqrt{2\omega_\pi(\vek_1)\,
2\omega_\pi(\vek_2)\, V_3^2}}\,, \label{39}
\end{equation}
while
$J_{nn_2n_3} (\vep)$ is the pionless overlap integral (\ref{2}), where $\vek
\equiv 0$, and $\bar y_{123}^{(h)} \to \bar y_{123} =
M_\omega\,\frac{i}{\omega_q} (q_i - \frac{p_i\,\omega_q}{2\,(\omega_q +
\omega_Q)}).$

Looking at (\ref{37}), one can realize, that it can be written as
\begin{equation} w^{(\pi\pi)}_{nm} (E) = ~^{(1)} w_{nm}^{(\pi\pi)}(E) -
~^{(2)}w_{nm}^{(\pi\pi)}(E)\label{40}
\end{equation}
and the first two terms of $~^{(1)} w_{nm}^{(\pi\pi)}(E)$ are
\begin{equation}
~^{(1)} w^{(\pi\pi)}_{nm}= \varphi(\vek_1)\, \zeta_{nm}(E')\,
\varphi(\vek_2)+ \varphi(\vek_2)\, \zeta_{nm}(E'')\,
\varphi(\vek_1)\,,\label{41}
\end{equation}
where $\zeta_{nm}$ depends on the energy $E'$ and $E''$ of the
$(Q\bar{Q})\,\pi_2$ and $(Q\bar{Q})\,\pi_1$ systems respectively,
its Lorenz invariant definition see below.
%%%%%%%%%%%%%%%%%%%%%%%%%%%%%%%%%%%%%%%
%%%%%%%%%%%%%%%%%%%%%%%%%%%%%%%%%%%%%%
\begin{figure}[t]
  \center{\includegraphics[angle=0,width=0.85\textwidth]{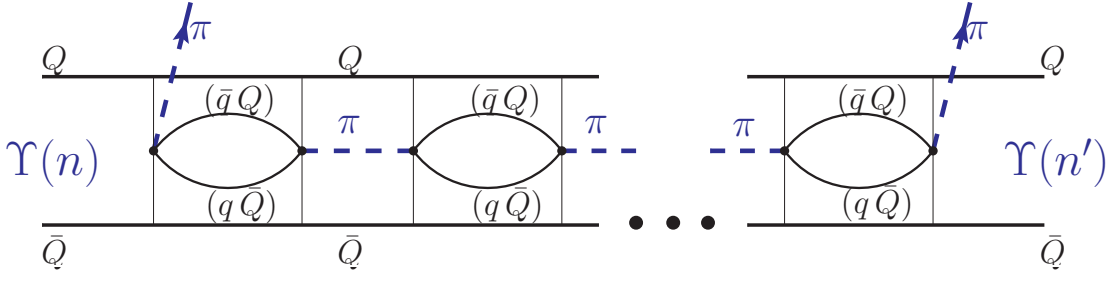}
    \caption{Rescattering series yielding a possible pole structure in $(\Upsilon(n')\pi)$. (Contribution to $a_{nm}$)}\label{fig:diagram3}}
\end{figure}
It is  clear that these terms are the first terms of the whole
rescattering series, depicted in Fig.\ref{fig:diagram3}, which can
be summed up as follows
\begin{equation}
~^{(1)} w^{(\pi\pi)}_{nm} \to ~^{(1)}W_{nm} = \varphi (\vek_1)
\left( \zeta\, \frac{1}{1-K\zeta}\right)_{nm} \varphi(\vek_2)+ \varphi(\vek_2)
\left( \zeta\, \frac{1}{1-K\zeta}\right)_{nm}
\varphi(\vek_1).\label{42}\end{equation}
Note, that for the transition $\Upsilon(n) \to \Upsilon(n')\,
\pi\pi$, $\zeta(E')$ and $K(E')$ in the first term in (\ref{42})
depend on the invariant mass $M^{(1)}_{inv}$ of
$\Upsilon(n')\,\pi_2$, while in the last term on the r.h.s. of
(\ref{42}), $\zeta(E^{\prime\prime})$ and $K(E^{\prime\prime})$
depend on the invariant mass $M^{(2)}_{inv}$ of
$\Upsilon(n')\,\pi_1$,
$$ M_{inv}^{(i)}  = \sqrt{ M_n^2-2\,M_n \omega_i + m^2_\pi},~~
i=1,2$$

\begin{figure}[t]
  \center{\includegraphics[angle=0,width=0.75\textwidth]{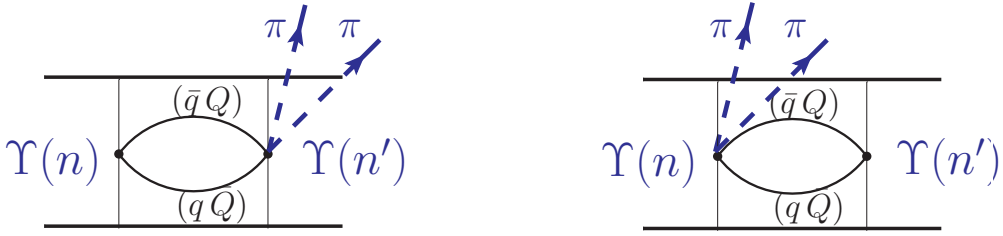}
  \caption{First order (in $\zeta)$ contributions to the factor $b_{nm} $ in Eq. (\ref{43}).\label{fig:diagram4}}}
\end{figure}

\begin{figure}[t]
  \center{\includegraphics[angle=0,width=0.85\textwidth]{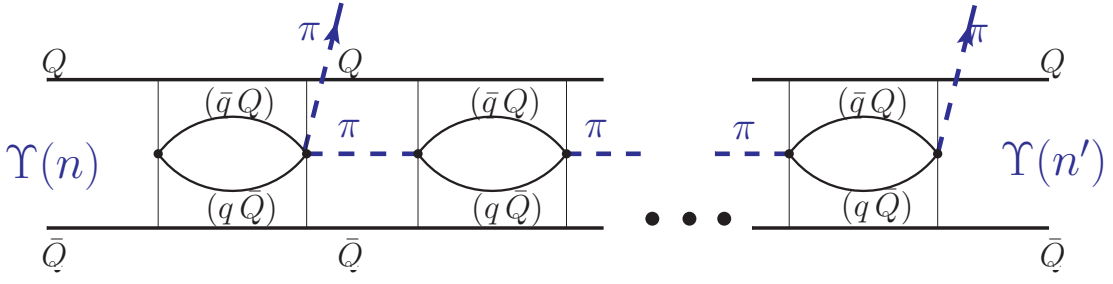}
  \caption{Rescattering series including double pion production vertex. (Contributing to
    $b_{nm}$).\label{fig:diagram5}}}
\end{figure}

We now turn to the last two terms in (\ref{37}), which contain two-pion vertex,
shown in  Fig.\ref{fig:diagram4}, and take into account, that the dependence on
$\vek_1, \vek_2$ there is contained in the factor $\varphi(\vek_1+\vek_2)$, as
was shown in \cite{13*,13**}, as well as in the series shown in
Fig.\ref{fig:diagram5}  (and the similar one with $\vek_1\leftrightarrow
\vek_2$). Hence the total amplitude of $(n, m)$ transition with emission of two
pions can be written as
\begin{equation}
w^{(\pi\pi)}_{nm} \equiv\Mc_{nm} =\varphi (\vek_1)\,\varphi(\vek_2)\, a_{nm} -
\varphi\,(\vek_1 +\vek_2)\,b_{nm}\, \label{43}
\end{equation}
where
\begin{eqnarray}
a_{nm} &=& \left( \zeta (E')\, \frac{1}{1- K(E')\,\zeta
(E')}\right)_{nm} + (E'\to E^{\prime\prime})_{nm}\nonumber\\
b_{nm}&=&\left(\zeta (E')\, \left(\frac{1}{1-K(E')\,\zeta(E')}+
(E'\to E'')\right)\right)_{nm} \label{44}
\end{eqnarray}
One can see, that (\ref{43}) satisfies Adler condition

\begin{equation}
a_{nm} (k_i =0) = b_{nm} (k_i=0), ~~\mathcal{M}_{nm}
(\vek_1=0)=0,\label{45}
\end{equation}
and for not very large $\vek_1,\, \vek_2$ one can write
approximately
\begin{equation}
\Mc_{nm} \approx {a_{nm}\, (\varphi(\vek_1)\,\varphi(\vek_2) -
\varphi(\vek_1+\vek_2))}.
\label{46}
\end{equation}

The probability of transition $(Q\bar Q)_n \to (Q\bar Q)_{n'}\,\pi\pi$ is
\begin{equation} dw((n) \to (n')\, \pi\pi) = |\Mc_{nn'}|^2\, \frac{d^3\vek_1}{(2\pi)^3}\frac{
d^3\vek_2}{(2\pi)^3}\, \frac{\pi}{2\,\omega_1\,\omega_2}\,\delta
(E_{n'}+\omega_\pi (\vek_1)+ \omega_\pi(\vek_2)- E_n) \label{47}\end{equation}
and the dipion decay width is
\begin{equation} \Gamma^{(\pi\pi)}_{nn'} = \int dw ((n)\to (n') \pi\pi) = \int
d \Phi\,|\mathcal{M}_{nn'}|^2\label{15ff}\end{equation}
where $d\Phi$ is the phase space factor
\begin{equation} d\Phi= \frac{1}{32\,\pi^3\, } \frac{(M_n^2+ M_{n'}^2
-q^2)(M_{n}+M_{n'})}{4\,M_{n}^3}\,\sqrt{(\Delta M)^2 -q^2}\,
\sqrt{q^2-4m^2_\pi}\,\, dq\,d\cos \theta\label{16f}\end{equation}
with the notations
%M= M[(Q\bar Q)_n], ~~ M' =M[(Q\bar Q)_{n'}],~~
$$  \Delta  M=M_{n}-M_{n'}$$
\begin{equation}
q^2\equiv M^2_{\pi\pi}= (k_1+ k_2)^2 = (\omega_1+\omega_2)^2 - (\vek_1
+\vek_2)^2. \label{17f}
\end{equation}

Finally, one can also study the process $e^+ e^- \to (Q\bar Q)_{n'}\, \pi\pi$,
with the amplitude
\begin{equation}
A_{n'}(E) =\sum_{m,n}\,c_m\left(\frac{1}{\hat E +\hat w-E}\right)_{mn}
\Mc_{nn'} \label{49}
\end{equation}
where $\Mc_{nn'}$ is given in (\ref{43}) and $c_m =\frac{4\,
\pi\,\alpha\, e_Q \sqrt{6} }{E^2} \psi_m (0)$ so that the
contribution $\Delta R$ of $(Q\bar Q)_{n'}\,\pi\pi$ to the
hadronic ratio $R$ is
\begin{equation} \Delta R ((n')\, \pi\pi) = \frac{72\, \pi\, e^2_Q}{E^2}\,
\left| \sum_{n,m} \psi_n (0) \left(\frac{1}{\hat E+\hat w-E}\right)_{nm}
\Mc_{mn'} \right|^2 d \Phi\label{50}\end{equation}

We now turn to an example of possible pionic bottomonium state in the reaction
$\Upsilon (5S) \to (\Upsilon (n')\, \pi)\,\pi$, which can proceed through the
chains
\begin{equation}
\Upsilon (5S) \to \sum_{n_2n_3} (B\bar B) _{n_2n_3}\, \pi \to \sum_{n'}
(\Upsilon(n')\, \pi )\,\pi. \label{51}
\end{equation}
We are interested in possible poles in the $J^P=1^+$ channel of
connected $(B\bar B)_{n_2n_3}$ and $\Upsilon(n')\,\pi$ states,
which are given by the equation
\begin{equation}
\det \Big[\delta_{nn'} - K_n (E)\,\zeta_{nn'}(E)\Big] =0. \label{52}
\end{equation}
Neglecting first nondiagonal elements of $\zeta_{nn'}$, one has an equation for
$E$ \begin{equation} K_n(E)\,\zeta_{nn}(E) =\frac{1}{N_c}\,\sum_{n_2n_3} \int
\frac{d^3\vep}{(2\pi)^3}\,
\frac{\chi^2_{nn_2n_3}(\vep)}{E_{n_2}(\vep)+E_{n_3}(\vep)-E} \int
\frac{d^3\vek}{(2\pi)^3}\frac{\varphi^2 (\vek)}{2\omega_\pi (k)\, (E_n (\vek)+
\omega_\pi(\vek) -E)}=1\label{53}\end{equation}
One can easily recognize in (\ref{53}) the norm of the kernel of the integral
equation (\ref{8}).

The analysis of (\ref{53}) starts with the list of thresholds in
$(B\bar B)_{n_2n_3}$ and in $\Upsilon(n')\,\pi$ channels in Table
\ref{tab.Thresh}.
\begin{table}[t]
\caption{List of thresholds (in GeV).  \label{tab.Thresh}}
\begin{center}
\begin{tabular}{cccccl}
  \hline  \hline
  & Threshold & $E_{th}$ &  & Threshold & $E_{th}$
  \\ \hline
  & $BB^*$ & 10.605 &  & $\Upsilon(1S)\,\pi$ & 9.60 \\
  & $B^*B^*$ & 10.650 &  & $\Upsilon(2S)\,\pi$ & 10.16 \\
  & $B_sB_s^*$ & 10.780 &  & $\Upsilon(3S)\,\pi$ & 10.495 \\
  & $B_s^*B_s^*$ & 10.830 &  & $\Upsilon(4S)\,\pi$ & 10.720\\
  &  &  &                    & $\Upsilon(5S)\,\pi$ &11.00 \\
  \hline\hline
\end{tabular}\end{center}
\end{table}
One can see, that the most important combination is $\Upsilon (3S)\, \pi
\leftrightarrow BB^*,\, B^*B^*$ with additional contribution of
$\Upsilon(2S)\,\pi$, $\Upsilon (1S)\,\pi$  and $B_sB^*_s,\, B^*_s B_s^*$ as a
next step. Since the maximal energy of our systems in the reaction (\ref{51})
is $M(5S) -m_{\pi}=10.725$ GeV, one is mostly interested in the energy region
9.78 GeV $\leq E \leq 10.72$ GeV.

%\newpage

%\begin{figure}
%\includegraphics[width=80mm,keepaspectratio=true]{zeta33.eps}
%\includegraphics[width=80mm,keepaspectratio=true]{K3.eps}
% \caption{a) Real  part (solid line) and imaginary part (broken line)
% of $\zeta_{33}$ for  channel $3\equiv \Upsilon(3S)\,\pi$. One can see sharp peaks at the thresholds $BB^*$ and
% $B^*B^*$; b) Real part of $K_3$ (solid line) and imaginary part  of $K_3(E)$
%(broken line). One can see a peak near the threshold of $\Upsilon
%(3S)\,\pi$.}
%\end{figure}

\begin{figure}[t]
\includegraphics[height=6.0cm]{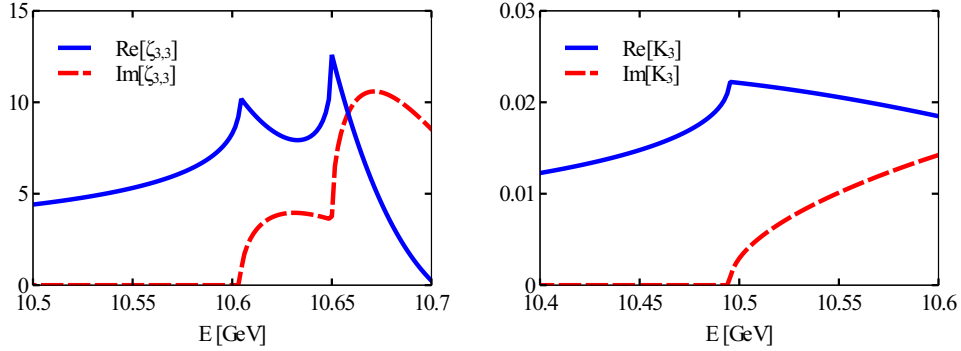}
\caption{Real (solid line) and imaginary (dashed line) parts of $\zeta_{33}(E)$
and $K_3(E)$ (see Eqs.(\ref{34}, \ref{35})). One can see cusp structures at the
thresholds $BB^*$, $B^*B^*$ and $\Upsilon (3S)\,\pi$
correspondingly.\label{fig:Zeta33 and K3}}
\end{figure}

\section{Results for  $\Upsilon (5S) \to \Upsilon (n'S)\,\pi\pi$}

In this section we study  numerical results for the reaction $\Upsilon(5S) \to
\Upsilon(n'S)\,\pi\pi$, and we shall be interested mainly in the possible
appearance of resonance-like structures in the systems $\Upsilon (n'S)\,\pi$,
$n'=1,2,3$. The resulting equations for  differential  and total probabilities
are given in  Eqs. (\ref{44}), (\ref{46}), (\ref{47}), (\ref{15ff}). The
coefficients $\bar y^{(\pi)}_{123}$ and parameters of $\Upsilon (nS)$ and $B,\,
B^*$ wave functions, needed for calculation of $\zeta_{nm}$ are given in
Eqs.(\ref{29}, \ref{30}) and Appendices 1 and 2.

It was assumed above, that the knowledge of wave functions and
channel coupling constant $M_\omega$ (one for all types of strong
decays) can describe all $CC$ phenomena and, in particular, level
shifts due to the $CC$. However, at this point one encounters the
fundamental difficulty, which was  studied in \cite{Geiger:1989yc}
and is not still resolved. The point is, that assuming a constant
$CC$, not depending on participants of decay process, one obtains
huge shifts of energy levels (several hundreds of MeV) due to
mixing with higher states. (To improve situation, in \cite{13} a
cut-off coefficient $\kappa \approx 0.5$ was introduced for
contribution of higher levels). This calls for a  detailed
scrutiny of our matrix elements $J^{(h)}_{nn_2n_3}(\vep, \vek)$
and basic matrix elements (without hadron emission)
$J_{nn_2n_3}(\vep)$, entering in the expressions for energy
shifts, see \cite{12} for details. Indeed in Eq. (\ref{2}) one can
see, that the string in the original hadron $n$, placed between
points $\veu, \vev$ is decaying at point $\vex$ into two hadrons,
placed between $\veu, \vex$ , and $\vex, \vev$ respectively. It is
clear, that when the point $\vex$ is far  away from the center of
the string $\veu, \vev$, the string breaking process does not
occur. Replacing distances between points by typical radii $R_n$,
$R_{n_2},\, R_{n_3}$ of states $n,n_2,n_3$ one obtains condition
(for $R_{n_2}\approx R_{n_3}$) $R^2_{n_2} \la \frac14 R^2_n
+\rho^2$, where $\rho$ is the typical string width, $\rho\approx
2\lambda\approx 0.3$ fm \cite{Cardoso:2011cs,Kuzmenko2004}. The
corresponding factor can be rigorously deduced in the formalism of
\cite{22}, and the resulting cut-off strongly decreases the $CC$
between states with incomparable sizes.  This is taken into
account below by assuming different $M_\omega$ for different decay
channels. We take for $M_\omega$ the value $1.0$ GeV in case of
$n=4,5$. For $n =1,2,3$ we take $M_\omega =0.1,\,0.2$ and $0.3$
GeV respectively, to take into account decay mismatch between the
sizes of $\Upsilon(nS)$ and $B\bar B^*$ systems, since $R(1S)
=0.2$ fm and $R(2S)= 0.4$ fm, while $R(B)\approx R(B^*)\approx
0.5$ fm. This mismatch is not taken into account for simplicity
reasons in the general definition of the overlap integral
(\ref{2}).

\begin{figure}
\includegraphics[height=6.5cm]{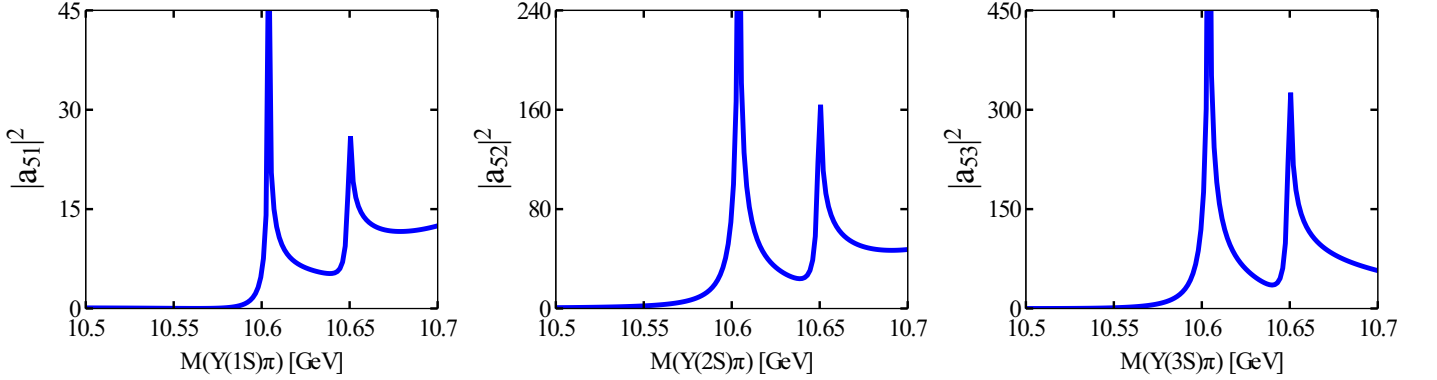}
\caption{Modulus squared of the first term in (\ref{44}) $|a_{nn'}|^2$ as a
function of the invariant mass of $(\Upsilon(n'S)\pi)$, computed for the
reaction $\Upsilon(5S) \to \Upsilon(n'S)\,\pi\pi$, $n'=1,2,3$.\label{fig:anm}}
\end{figure}

In the beginning we have estimated $K_n(E)$ and $\zeta_{nm} (E)$
in (\ref{53}) for $n,m=1,2,3,4,5$, using parameters of SHO
functions, which were used before in \cite{13**,13***}, they are
given in the Appendix 2. Real and imaginary parts of $\zeta_{33}
(E)$ and $K_3 (E)$ are given in Fig.\ref{fig:Zeta33 and K3}. The
possibility of peaks e.g. in $\Upsilon(n'S)\,\pi$ system is
demonstrated in Fig.\ref{fig:anm},  where we plot the quantity
$\left|a_{5n'}\right|^2$ for $n'=1,2,3$ respectively  (or more
exactly the first term  of (\ref{44})) as function of the
invariant mass $M^{(1)}_{inv}$. One can see sharp peaks near the
thresholds $BB^*, B^*B^*$ at 10.6 and 10.65 GeV respectively. In
the total distributions, however, a more complicated combination
of terms enters, as seen from (\ref{44}), and one should calculate
a symmetrized  in $M^{(1)}_{inv},\, M^{(2)}_{inv}$ expression, and
moreover take into account appropriate phase space factor. One can
associate these peaks with poles, situated in the vicinity of
these thresholds (see discussion in Appendix 3).

 Note, that  two
peaks in  Fig.\ref{fig:anm} occur from combination $\Upsilon
(n'S)\pi_1$. However in the full decay distribution $dw((5)\to
(n') \pi\pi)$ the symmetrized sum (\ref{44}) enters, which
produces an additional mirror reflected pair of peaks in $dw$ if
considered as function of $M_{inv}^{(1)},\, q$, since

\begin{equation}
\left(M^{(1)}_{inv}\right)^2+ \left(M^{(2)}_{inv}\right)^2 =
M_{n}^2 + M_{n'}^2 + 2\, m^2_\pi - q^2.\label{56}
\end{equation}

\begin{figure}
\includegraphics[height=6.5cm]{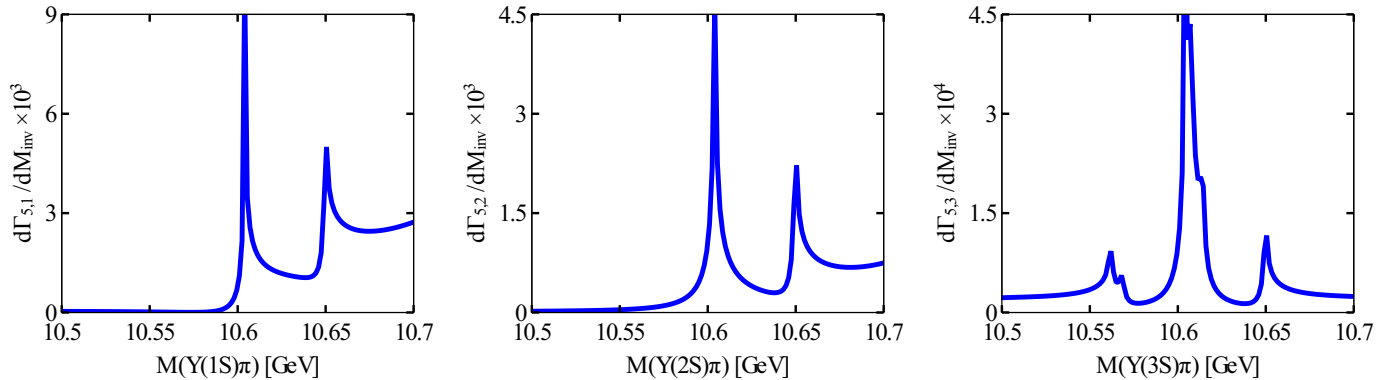}
\caption{The distribution $d\Gamma_{nn'}/dM_{inv}$ for the transition
$\Upsilon(nS)\rightarrow \Upsilon(n'S)\,\pi\pi$ as a function of the invariant
mass of $(\Upsilon(n'S)\pi)$ for $n'=1,2,3$.\label{fig:Gammanm}}
\end{figure}

The full probability distribution (\ref{47}) contains symmetric
sum of two rescattering series for $\Upsilon(n'S)\,\pi_1$ and
$\Upsilon(n'S)\,\pi_2$ respectively. When plotted as function of
$M_{inv}^{(1)}$ it contains both poles of the first series at
$M_{inv}^{(1)}=10.610$ and $10.650$ GeV, and also poles from the
second series at the points
$M_{inv}^{(1)}(M_{inv}^{(2)}=10.610,\,10.650)$ defined by
Eq.(\ref{56}). In this way one obtains Fig.\ref{fig:Gammanm},
where for the case $(n,n')=(5,1)$ and $(5,2)$ the secondary poles
are out of mass interval, while for the case of $(5,3)$ the pair
of secondary poles overlap with the proper poles of
$M_{inv}^{(1)}$. In experiment one can separate points on Dalitz
plot relating to $M_{inv}^{(1)}$ and $M_{inv}^{(2)}$, which
results in two plots with the same pair of peaks.

Recently experimental data on the reaction  $\Upsilon (5S) \to
\Upsilon(n'S)\,\pi\pi,\, n'=1,2,3$ appeared in \cite{23}, and the
experimental distributions $d w ((5)\to (n') \pi\pi)$  are
presented in Fig.5 of \cite{23} as functions of $\min
\{M^{(1)}_{inv}, M^{(2)}_{inv}\}$ and $\max \{M^{(1)}_{inv},
M^{(2)}_{inv}\}$. In order to compare our results with
experimental data, we calculate distribution (\ref{47}) in terms
of invariant masses of $\Upsilon(n'S)\,\pi$ and $\pi \pi$ systems.
We use expressions (\ref{46}), (\ref{16f}) and formulas from
\cite{13*} to express quantities like $\omega_\pi(\vek_1),\,
(\vek_1+\vek_2)^2$ etc. via variables $(q,\, M_{inv})$. After
integration over $q$ we obtain the distribution in terms of
$M_{inv}(\Upsilon(n'S)\,\pi)$. The result is presented on
Fig.\ref{fig:Gammanm}.
Comparing experimental data (left and right plots, glued together
for $n'=3$ and left plots for $n'=1,2$ on Fig.5 from \cite{23})
with our theoretical calculations, one can see a close similarity
in the general form and position of $Z_b(10610)$ and $Z_b(10650)$
peaks, which in both cases appear at $BB^*,\, B^*B^*$ thresholds
for all available $n'=1,2,3$. We do not intend to reproduce here
all features of experimental data, which depend strongly on
details of wave function profile, but we study mostly the dynamics
of process. For the better description we only slightly changed
w.f. of $\Upsilon(nS)$, presented in Appendix 2, by $\sim20 \%$. A
more detailed quantitative comparison of our data with experiment
\cite{23}, and decay distributions as functions of $q \equiv
M_{\pi\pi}$ and $\cos \theta$, are now in progress and will be
reported elsewhere.

\section{Conclusions and  outlook}

Summarizing our results, we shall  stress the main features of our approach. We
have considered the interaction of a light hadron with heavy quarkonium,
arising solely due to  transition to intermediate states of two heavy-light
mesons. No direct interaction between light hadron and heavy quarkonium or else
between two heavy-light mesons is assumed, therefore our dynamics has nothing
to do with molecular states in the strict sense. Our mechanism is also distinct
from dynamics of $(4q)$ or hybrid states.

We have constructed explicitly transition vertex for the strong decay with
emission of a light hadron from the first principles, and then all dynamics is
defined by the overlap integrals of all wave functions involved, i.e. heavy
quarkonia and heavy-light mesons. The latter have been found in \cite{22*} from
the relativistic Hamiltonian, containing only first principle input, and
accuracy of Gaussian representation was checked in \cite{12,16}.
%Thus our whole
%construction contains till now no fitting parameters and one can check with
%this whether it gives interaction strong enough to produce resonance poles.

The answer given in Figs.\ref{fig:anm},\ref{fig:Gammanm} is positive for the
case of pion-bottomonium system, and agrees with the recent experimental data
\cite{23} at least qualitatively. More detailed calculations are still needed
to check all results quantitatively $vs$ experiment and  this program is now
under study. Other systems should be treated as well, e.g. $\pi\, h_b$, studied
in \cite{23}, and the series of $Z(4430),\,Z_1(4050),\, Z_2 (4250)$ resonances
in the $(c\bar c)\pi$ system. All formalism used above for bottomonium can be
applied to the charmonium case without modifications; for the vector charmonium
or bottomonium ($h_b,\,\chi_b,\,h_c,\,\chi_c)$ one can use transition vertices,
given in Appendices 1,2, and symmetry properties of the whole amplitude will be
different. This work is now in progress.

One should stress at this point, that resonances, found above in
$\Upsilon(n'S)\pi$ system are specifically multichannel ones, in the sense,
that they belong (and appear) in all $n'$ channels and are given by zeros of
$\det[1-\hat K\hat\zeta]$.

At this point it is convenient to compare predictions of molecular and $CC$
models for the positions of resonances. As one can see in Fig.\ref{fig:anm},
our $CC$ model predicts sharp peaks at $\bar B B^*, \bar B^* B^*$ thresholds
(possibly due to virtual multichannel poles) in all $\Upsilon(n'S)\pi$
$(n'=1,2,3)$ channels, and this effect is due to combined contribution of $\hat
\zeta$ and $\hat K$ terms, i.e. pure $CC$ effect. In contrast to that,  the
pure molecular picture, when poles in the $B B^*,\, B^* B^*$ systems appear due
to  direct internal interaction (i.e. without  $\pi\Upsilon (nS)$  channels),
the poles appear in all $\zeta_{nn}(E)$ terms near corresponding thresholds.
This can be seen simply in the definition of  $\hat \zeta(E)$ in (\ref{35}),
where $\chi_n \chi_{n'}$ is multiplied by the free $BB^* \,(B^*B^*)$ Green's
function. In case of strong interaction in $BB^*$, $B^*B^*$ systems, this Green
function is replaced by the exact one and should contain pure molecular poles
i.e. $\zeta_{nm} (E) = \frac{C_n C_m}{E_0-E}$. Insertion of this form into the
$ \det [1-\hat K \hat \zeta] =0$ yields the $n$-th order equation for roots in
energy, and these $n$ roots are all strongly displaced from the original places
at the thresholds for large $K\zeta$ coupling (large $C_n$), and are almost
degenerate $n$ poles at thresholds for small coupling. Both pictures are
different from the experimental data -- the same two poles at $ \bar B B^*$ and
$\bar B^* B^*$ in all $\pi \Upsilon (nS)$ and $\pi h_b(mP)$ channels. Thus the
situation with the only pole at each threshold is possible and characteristic
for our multichannel $CC$ resonance and is unlikely for pure molecular states.

Another property is, that resonances appear most likely, when thresholds in
$\hat K_n$ and $\hat \zeta_{nn} $ for some $n$ are close to each other, and
then this channel $n$ will be the dominant one, making $\det[1-\hat
K\hat\zeta]$ close to zero. In the case of $\pi \Upsilon (n'S)$ the dominant
channels are those with $n'=3,4$, as can be seen from Table in section 4.
Therefore the visible channel $h\,(Q\bar Q)_{n'}$,where a peak is found, is not
necessarily the dominant one, as might be in the  case of $Z(4430)$ with a peak
seen in $\pi^+\psi'$ channel. The dynamical reason, why proximity of thresholds
is favorable for the appearance of a resonance, is that both $\textrm{Re}[K_n
(E)]$ and $\textrm{Re}\,[\zeta_{nn}(E)]$ are decreasing fast enough away from
thresholds, making their product maximal for the coinciding thresholds.

In all discussion above the notion of resonances (or virtual and
real poles) was stressed, and hence the  whole sum of rescattering
series as in Figs.\ref{fig:diagram3},\ref{fig:diagram5}, were
considered. But it is possible, that already the first terms of
this series can contribute to enhanced correlations, which look
like bumps in decay distributions. This approach was  considered
in two recent papers \cite{25, 26}, and is especially important in
the case, when high spins and angular momenta are involved.
Relation of this approach to our methods of present paper seems to
be practically important and the corresponding work is planned for
the future.

Recently several papers appeared \cite{32,33}, where
$Z_{b}(10610)$ and $Z_{b}(10650)$ are considered as molecular
states and treated in the QCD sum rule method \cite{32}, while  in
\cite{33} these states were  originally supported to be
$\chi_{b_1}$ and $\chi_{b_1}'$ shifted to  the  $BB^*$ and
$B^*B^*$ thresholds (as it happens in $X(3872)$ case). As was
argued above in both cases the poles produced appear in
$\zeta_{nn'}$ and would be strongly shifted in the rescattering
series of Fig.\ref{fig:diagram1}, yielding 5 peaks at different
masses in $\Upsilon(nS)\,\pi$, $h_b(mP)\,\pi, ~~ n=1,2,3; ~~
m=1,2$.

In \cite{34} an interesting analysis of $\Upsilon (5S) \to \Upsilon(2S)\,
\pi\pi$ decay is presented, demonstrating important contribution of
$Z_{b}(10610),\,Z_{b}(10650)$ to the decay distribution in $M(\pi^+\pi^-)$ and
$\cos\theta$  a detailed comparison of these results with our  approach and
previous results in \cite{13***} is now in progress.

The authors are  grateful to A.M.Badalian, R.Mizuk and P.N.Pakhlov  for many
useful discussions. The financial support of Dynasty Foundation to V.D.O. and
RFBR grant no.09-02-00 620a is gratefully acknowledged.

\vspace{2cm}

%\newpage

{\bf Appendix 1 }\\
{\bf Calculation of the transition kernel $\bar y^{(h)}_{123}$}
\setcounter{equation}{0} \def\theequation{A1.\arabic{equation}} \vspace{1cm}

\begin{figure}[t]
\center{
\includegraphics[angle=0,width=0.35\textwidth]{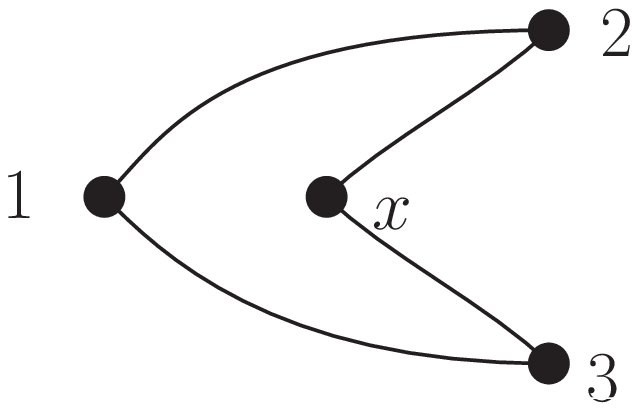}
\includegraphics[angle=0,width=0.35\textwidth]{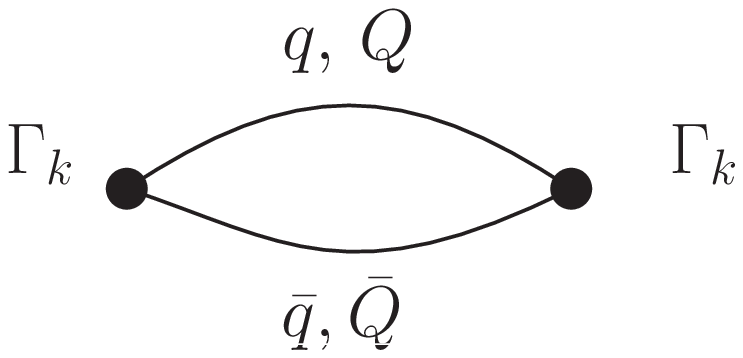}
\caption{Decay matrix vertices.\label{fig:Verticies}}}
\end{figure}

We use here, as well as in  \cite{12,13*,13**} the spin-tensor representation
of wave functions and $Z$ factors instead of technic of Clebsch-Gordan
coefficients. We start with the $(4\times  4)$ fully  relativistic technic
given in \cite{13*,13**}, where $\bar y_{123}$ can be written through the
so-called $Z$ factors (we omit the superscript $h$ for time being, cf.
Eq.(A2.24) of \cite{13*}) \begin{equation} \bar y_{123} =
\frac{Z_{123x}}{\sqrt{\Pi_{i=1,2,3}Z_i}}\label{A1.1}\end{equation} where
$Z_{123x}$ and $Z_i$ are $Z$ factors for the total process and for individual
hadrons respectively participating in the transition process $1\to 23 $ or
$1\to 23\, h$, shown in Fig.\ref{fig:Verticies}. Defining projection factors
for quarks and antiquarks $\Lambda_k^\pm$, where subscript $k$ refers to light
quarks, $ k=q$, or heavy quarks, $k=Q$, one can write
\begin{equation} Z_{123x} =tr (\Gamma_1\, \Lambda^+_Q\, \Gamma_2\,
\Lambda^-_q\, \Gamma_x\, \Lambda_q^+\, \Gamma_3\, \Lambda_Q^-),
\label{A1.2}\end{equation}
\begin{equation} Z_i = tr (\Gamma_i\, \Lambda^+_k\,
\Gamma_i\, \Lambda_k^-),\label{A1.3}\end{equation} and
\begin{equation} \Lambda_k^{\pm} = \frac{m_k \pm \omega_k \gamma_4
\mp i p_i^{(k)} \gamma_i}{2\omega_k} , ~~ k=q, Q.\label{A1.4}\end{equation}
Note, that the sum over (i) in (A1.4) is for $i=1,2,3,$ also in the c.m. system
$p_i^{(1)} = - p_i^{(2)} \equiv p_i$, while $m_k$ is the current quark mass and
$\omega_k$ is the average kinetic energy of quark $k$ in the hadron.

The operators $\Gamma_i$ correspond to quantum numbers of a hadron, and are
given in Table 1 below, while $\Gamma_x$ refers to the process under
investigation, for the case when no hadron is emitted, $\Gamma_x=1$, while for
the pion emission $\Gamma_x^{(\pi)}= \gamma_5$ and for   vector particles
$\Gamma_x^{(v)} = \gamma_i$.
%\newpage

%%%%%%%%%%%%%%%%%%%%%%%%%%%%%%%%%%%%%%%%%%%%%%%%%%%%%%%%%%%%%%%%%%%%%%%%%%%%%%
\begin{table}%[h]
\caption{\label{tab.5} Bilinear operators $ \bar \psi\, \Gamma_i\, \psi$ and
their $(2\times 2)$ forms (Notations see in the text).}
\begin{center}
\begin{tabular}{ccccc}
\hline \hline $J^{PC}$& $~^{2S+1}L_J$& $\Gamma_i$ &$ (2\times 2)$ form& $\Gamma_{\rm red}$\\
\hline
$0^{-+}$& $~^1S_0$& $-i\gamma_5$& $\tilde v^cv-\tilde w^c w$&$\frac{1}{\sqrt{2}}$\\

$1^{--}$& $~^3S_1$&$\gamma_ i$& $-(\tilde v^c\sigma_i v+\tilde w^c \sigma_iw)$&$\frac{1}{\sqrt{2}}\sigma_i$\\
$1^{+-}$& $~^1P_1$& $-i\gamma_5 \stackrel{\leftrightarrow}{\partial}_i     $&
$\tilde v^c\stackrel{\leftrightarrow}{\partial}_iv-
\tilde w^c\stackrel{\leftrightarrow}{\partial}_i w$&$\sqrt{\frac32}n_i$\\

$0^{++}$& $~^3P_0$& 1& $i (\tilde v^cw-\tilde w^c v)$&$\frac{1}{\sqrt{2}}\vesig\ven$\\

$1^{++}$& $~^3P_1$&$\gamma_i\gamma_5$& $-(\tilde v^c\sigma_i w+\tilde w^c \sigma_iv)$&$\frac{\sqrt{3}}{2} e_{ikl}\sigma_k n_l$\\
$2^{++}$& $~^3P_2$& $\gamma_i\stackrel{\leftrightarrow}{\partial}_k + \gamma_k
\stackrel{\leftrightarrow}{\partial}_i-\frac23\, \delta_{ik}\hat
\partial$ & $-(\tilde v^c\rho_{ik} v+ \tilde w^c\rho_{ik}
w)$&$P_2(\vesig,\ven)$\\

$2^{-+}$& $~^1D_2$& $(\stackrel{\leftrightarrow}{\partial}_i
 \stackrel{\leftrightarrow}{\partial}_k-\frac13\,
\delta_{ik}(\stackrel{\leftrightarrow}{\partial})^2)\gamma_5 $ & $i (\tilde
v^c\omega_{ik}w- \tilde w^c\omega_{ik}
v)$&-\\
$2^{--}$& $~^3D_2$& $(\gamma_i \stackrel{\leftrightarrow}{\partial}_k +
\gamma_k \stackrel{\leftrightarrow}{\partial}_i-\frac23\, \delta_{ik}\hat
\partial)\gamma_5$ & $-(\tilde v^c\rho_{ik} w+ \tilde w^c\rho_{ik}
v)$&-\\ $1^{--}$& $~^3D_1$& $\gamma_i \omega_{ik}$&  $-( v^c\sigma_i
\omega_{ik}v+ \tilde w^c\sigma_i \omega_{ik}
w)$&$\frac32 \sigma_i (n_i n_k -\frac13 \delta_{ik})$\\
\hline\hline
\end{tabular}
\end{center}
Here $P_2(\vesig,\ven)\equiv \frac34 (\sigma_i n_l + \sigma_e n_i -\frac23\,
(\vesig \ven)\, \delta_{il})$.

\end{table}
%%%%%%%%%%%%%%%%%%%%%%%%%%%%%%%%%%%%%%%%%%%%%%%%%%%%%%%%%%%%%%%%%%%%%%%%%%%%%%%%%%%%%%%%%%%

%\begin{figure}[t]
%  \center{\includegraphics[angle=0,width=0.35\textwidth]{FigVerticies.eps}
%    \caption{Decay matrix vertices}}
%\end{figure}

%\begin{figure}[t]
%  \center{\includegraphics[angle=0,width=0.35\textwidth]{FigVerticies2.eps}
%    \caption{Decay matrix vertices}}
%\end{figure}

%%%%%%%%%%%%%%%%%%%%%%%%%%%%%%%%%

In this way one obtains\begin{equation} Z_{D,B} = \frac{\Omega-\omega}{\Omega},
~~ Z_{\psi, \Upsilon} = \frac{\frac43\, \Omega^2_{\psi, \Upsilon} + \frac23\,
m^2_{c,b}}{\Omega^2_{\psi,\Upsilon}},\label{A1.5}\end{equation} where
$\Omega_{D,B} = \lan \sqrt{\vep^2 + m^2_{c,b}}\ran_{D,B} , ~~
\Omega_{\psi,\Upsilon} =\lan \sqrt{\vep^2+m^2_{c,b}}\ran_{\psi, \Upsilon}$ are
calculated in \cite{13*} and given in Appendix 1 of [14].

Examples of relativistic ($4\times 4$) expressions of $\bar y_{123}$ are given
in \cite{13*,13**}, e.g. for $(1^{--})_n \to B B^*\pi\, (DD^*\pi)$ and
$(1^{--})_n \to B^* B^*\pi\, (D^*D^*\pi), ~ \bar y^{(\pi)}_{123}$ are given in
(\ref{29}),(\ref{30}) while for $(1^{--}) \to B B^*$ \begin{equation} \bar
y_{123^*} = \frac{i\,m_Q}{2\,\Omega^2\,\omega} \left( q_i\, (2\Omega+\omega) -
\pi \frac{\omega\Omega}{\omega +\Omega} \right)\approx
\frac{iq_i}{\omega}.\label{A1.5a}\end{equation}
% and for $(1^{--})_n \to B^* B^*$ \begin{equation} \bar y_{12^*3^*}= \end{equation}

We now turn to the case of $(2\times 2)$ formalism, introduced in \cite{12},
where resulting kernels are denoted as $\bar y^{\rm red}_{123}$, and are
computed  according to

\begin{equation} \bar y^{\rm red}_{123} = tr
\{\Gamma^{(n_1)}_{red}\,\Gamma^{(n_2)}_{red}\,(\vesig \veq)\,
\Gamma^{(n_3)}_{\rm red}\}\label{A1.7}\end{equation}
while for emission of an additional pion in $ \bar y^{(\pi){\rm~ red}}_{123}$
one should omit the factor $(\vesig \veq)$ in (\ref{A1.7}). In this way one
obtains (\ref{29}), (\ref{30}). Note, that normalization of states in ($2\times
2)$ formalism is different, and one should sum up over all polarizations in
initial and final states (extra factor of $1/\sqrt{2}$ is in (\ref{A1.7}) as
compared to (\ref{A1.5a}), and  $\frac{1}{\omega} \to \frac{1}{m+\varepsilon_n
+ \lan U\ran -\lan V\ran}$.

Of special interest are the transition kernels for $P$ states, where one has
for $~^3P_1\to (BB^*\pi), ~~ \bar y^{(\pi){\rm~ red}}_{123} =
\frac{\sqrt{3}}{2}\, e_{inl}\, n_l$ and for $~^1P_1 \to (B B^*\pi), ~~\bar
y^{(\pi){\rm~ red}}_{123} = 0$, while the $(4\times 4)$ formalism yields
\begin{equation} \bar y^{(\pi)}_{123} = \frac{\sqrt{3}}{4}\,
\frac{m_Q}{\Omega_Q}\, \frac{p_i}{p}\, e_{klm}\,(k-p)_n\, \left(1/2\,
p-q\right)_l.\label{A1.8}\end{equation}
For $~^1 P_1\to (B^* B^* \pi)$  one has $\bar y^{(\pi){\rm~ red}}_{123} =
\sqrt{\frac32}\,n_i\, \delta_{kl}$, and for $~^3P_1\to (B^*B^*\pi), ~~ \bar
y^{(\pi){\rm~ red}}_{123} = i\,\frac{\sqrt{3}}{2}\, n_l\, (\delta_{ls}
\delta_{ti} - \delta_{si} \delta_{tl})$  $i,s,t$ are indices of polarizations.
%\newpage

\vspace{1cm}
{\bf Appendix 2 }\\
{\bf Wavefunctions of  heavy quarkonia and heavy-light mesons }
\setcounter{equation}{0} \def\theequation{A2.\arabic{equation}} \vspace{0.5cm}

In Eq.(\ref{2}) $R_{Q\bar{Q}}^{(n_1)}$, $R_{Q\bar{q}}^{(n_2)}$ and
$R_{\bar{Q}q}^{(n_3)}$ are   series of oscillator wave functions, which are
fitted to realistic wave functions. We obtain them from the solution of the
Relativistic String Hamiltonian, described in \cite{22*}.

In position space the basic SHO radial wave function is given by
%(see e.g. the textbook by Flugge \cite{Flugge1971})
\begin{eqnarray}
&&R_{nl}^{\scriptscriptstyle
SHO}(\beta,\,r)=\beta^{3/2}\sqrt{\frac{2(n-1)!}{\Gamma(n+l+1/2)}}\,(\beta
r)^l\,e^{-\beta^2r^2/2}\,L_{n-1}^{l+1/2}(\beta^2r^2)\\
\nonumber &&\int\limits_0^{\infty}\,\big(R^{\scriptscriptstyle
SHO}_{nl}(\beta,\,r)\big)^2\,r^{\,2}\,dr=1
\end{eqnarray}
where $\beta$ is the SHO wave function parameter, and
$L_{n-1}^{l+1/2}(\beta^2r^2)$ is an associated Laguerre polynomial. The
realistic radial wave function can be represented as an expansion  in the full
set of oscillator radial functions:
\begin{equation}\label{}
    R_{n\,l}(r)=\sum_{k=1}^{k_{max}}c_k\,R_{kl}^{\scriptscriptstyle
SHO}(\beta,\,r).
\end{equation}
\begin{table}
\caption{\label{tab.betta}Effective values $\beta$ (in GeV) and coefficients
$c_k$ of the series of oscillator radial wave functions
$R_{kl}^{\scriptscriptstyle SHO}(\beta,\,r)$ which are fitted to realistic
radial wave functions $R_{nl}(r)$ of charmonium, bottomonium, B and D mesons.}
\begin{center}
\begin{tabular}{lclllll}
\hline\hline State  & $\beta$ & $c_1$ & $c_2$ & $c_3$ & $c_4$ & $c_5$\\
\hline

\multicolumn{7}{c}{Bottomonium}\\
1S & 1.27 & 0.977164 & -0.151779 & 0.141319 & -0.020857 & 0.036863 \\
2S & 0.88 & -0.18823 & 0.953901 & -0.135824 & 0.181277 & -0.001509 \\
3S & 0.76 & -0.128081 & -0.145885 & 0.936255 & -0.169962 & 0.226281 \\
4S & 0.64 & -0.019432 & -0.149876 & -0.362548 & 0.88647 & 0.014308 \\
5S & 0.6 & -0.011183 & -0.016911 & -0.182019 & -0.403138 & 0.853936 \\
1P & 0.93 & 0.977994 & -0.165514 & 0.122975 & -0.018631 & 0.024364 \\
2P & 0.76 & -0.092083 & 0.971982 & -0.137347 & 0.162174 & -0.002707 \\
%3P & 0.69 & -0.101792 & -0.07947 & 0.955806 & -0.156468 & 0.20532 \\
1D & 0.8 & 0.979042 & -0.168619 & 0.11132 & -0.016998 & 0.018383 \\
2D & 0.69 & -0.063135 & 0.979871 & -0.117356 & 0.145358 & 0.000973 \\
\multicolumn{7}{c}{Charmonium}\\
1S & 0.7 & 0.97796 & -0.169169 & 0.117682 & -0.019694 & 0.025113 \\
2S & 0.53 & -0.121144 & 0.973054 & -0.130808 & 0.141495 & 0.00097 \\
3S & 0.48 & -0.096897 & -0.086156 & 0.961504 & -0.155931 & 0.178935 \\
4S & 0.43 & -0.021639 & -0.128342 & -0.215657 & 0.947215 & -0.028308 \\
5S & 0.41 & -0.010701 & -0.022826 & -0.157891 &-0.258875& 0.925921\\
1P & 0.57 & 0.976869 & -0.184163 & 0.105506 & -0.018941 & 0.017215 \\
2P & 0.48 & -0.063059 & 0.981868 & -0.123012 & 0.127035 & 0.000588 \\
%3P & 0.45 & -0.078868 & -0.051711 & 0.971098 & -0.140116 & 0.164452 \\
1D & 0.51 & 0.979118 & -0.178313 & 0.095356 & -0.016095 & 0.013123 \\
2D & 0.45 & -0.044316 & 0.986084 & -0.107408 & 0.117002 & 0.001907 \\
\multicolumn{7}{c}{D meson}\\
1S     &0.48& \multicolumn{5}{c}{c=1}\\
\multicolumn{7}{c}{B meson}\\
1S     &0.49& \multicolumn{5}{c}{c=1}\\
\hline\hline
\end{tabular}
\end{center}
\end{table}
% FIGURE 1
%\begin{figure}[h]
%\begin{minipage}[h]{0.45\linewidth}
%\center{\includegraphics[angle=270,width=1.15\textwidth]{fig10.eps} \\
%a) $R^{(n_1)}_{Q\bar{Q}}(3S)/\sqrt{4\pi}$}
%\end{minipage}
%\hfill
%\begin{minipage}[h]{0.45\linewidth}
%\center{\includegraphics[angle=270,width=1.15\textwidth]{fig11.eps} \\
%b) $R^{(n_1)}_{Q\bar{Q}}(2P)/\sqrt{4\pi}$}
%\end{minipage}
% \caption{Realistic radial w.f. (divided by $\sqrt{4\pi}$) of charmonium
% $3S$ and $2P$ states (broken lines) and the series of oscillator functions with $k_{max}=5$ (solid lines).
% Note that the solid curves are  almost indistinguishable from the broken ones.}
%\label{ris:image1}
%\end{figure}
%% FIGURE 1
%
%
Effective values of oscillator parameters $\beta$ and coefficients $c_k$ are
obtained minimizing $\chi^2$ and listed in the Table
\ref{tab.betta}\footnote{Typos of the sign convection are corrected for the
$1S,2S,3S$ charmonium states of \cite{12}}. In the momentum space the SHO
radial wave function is given by \footnote{Note a typo in the equation for
$R_{nl}^{\scriptscriptstyle SHO}(\beta,\,p)$ of \cite{12}}
\begin{eqnarray} \nonumber
&&R_{nl}^{\scriptscriptstyle
SHO}(\beta,\,p)=\frac{(-1)^{n-1}\,(2\pi)^{3/2}}{\beta^{3/2}}\,\sqrt{\frac{2(n-1)!}{\Gamma(n+l+1/2)}}\,\left(\frac{p}{\beta}
\right)^l\,e^{-p^2/2\beta^2}\,L_{n-1}^{l+1/2}\left(\frac{p^2}{\beta^2}\right)\\
&&\int\limits_0^{\infty}\,\big(R^{\scriptscriptstyle
SHO}_{nl}(\beta,\,p)\big)^2\,\frac{p^{\,2}\,dp}{(2\pi)^3}=1
\end{eqnarray}

Then using  Eq. (A16) from Appendix A of \cite{13**}, one can write
$I_{nn_2n_3} (\vep)$ in (\ref{32}) from $n_2=n_3=1$ as
$(\beta_1=\beta_{Q\bar{Q}},\,\beta_2=\beta_{Q\bar{q}})$
\begin{equation} I_{n11} (\vep) = \tilde c_n\, (-1)^{n-1}\, 2\,
\frac{(2n-1)!}{(n-1)!}\, \Phi(-(n-1) ,~ \frac32, ~ \mathbf{f}^2)\,
\frac{y^{n-1}}{(2\sqrt{\pi})^3} \left( \frac{2\beta^2_1\beta^2_2}{\Delta_n }
\right)^{3/2}\label{A1.1}\end{equation}
where \begin{equation} \Delta_n  = 2 \beta^2_1 + \beta^2_2, ~~ y =
\frac{2\beta^2_1-\beta^2_2}{2\beta_1^2+ \beta^2_2}, ~~ \mathbf{f} =
\frac{2\,\vep\, \beta_1}{\Delta_n\,\sqrt{y}}, \label{A1.2}\end{equation}
\begin{equation} \tilde c_n = \left(\frac{2\pi}{\beta_1}\right)^{3/2}
\frac{(2\sqrt{\pi}/\beta_2)^3}{\sqrt{2^{2n} \pi^{3/2} (2n
-1)!}}.\label{A1.3}\end{equation}
Here $\Phi$ is the confluent hypergeometric series, $$\Phi(\alpha, \gamma,
x)=1+ \frac{\alpha}{\gamma \, 1!}\,x + \frac{\alpha(\alpha+1)}{\gamma
(\gamma+1)2!}\,x^2+...$$
%
%For $B$ and $B^*$ mesons the quality of the SHO approximation is
%given in Fig.4 of \cite{15}. The values of $\beta_2$ were found in
%\cite{13*}; for $B,B^*$ mesons, $\beta_2(B) =0.49$ GeV, for
%$D,D^*$ mesons from \cite{12} $\beta_2 (D) =0.48$ GeV. Also in
%Table V in \cite{12} are given $\beta_1$ and coefficients of the
%five terms in the series of SHO functions for states of charmonia
%$1S, 2S, 3S, 2P$. The values of $\beta_1$ and $c_k$ for bootmonia
%is presented on Table

\vspace{1cm}
{\bf Appendix 3}\\
{\bf Analytic structure of hadron-qurkonium resonances}
\setcounter{equation}{0} \def\theequation{A3.\arabic{equation}} \vspace{0.5cm}

\begin{figure}[t]
\includegraphics[height=4.0cm,angle=0,width=0.4\textwidth]{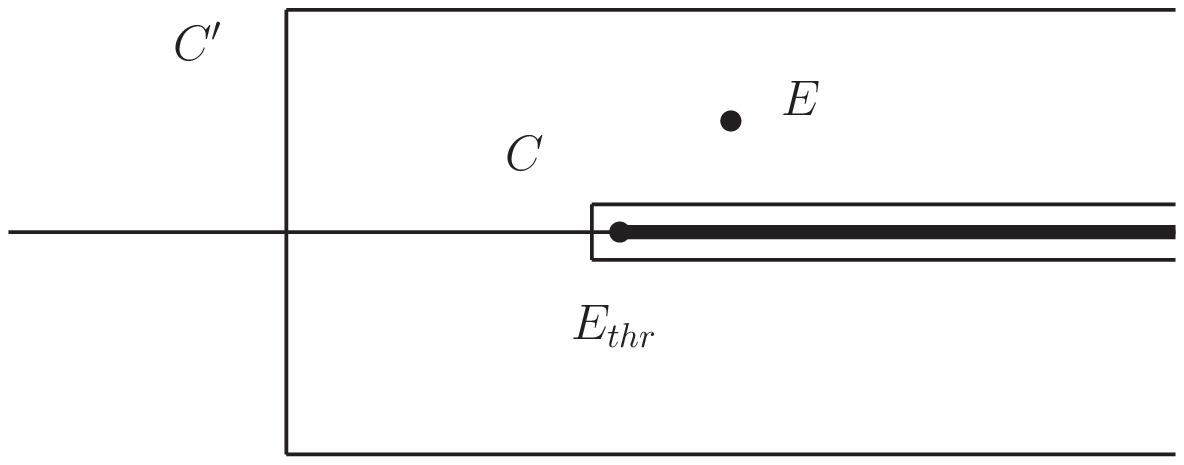}
\includegraphics[height=4.0cm,angle=0,width=0.3\textwidth]{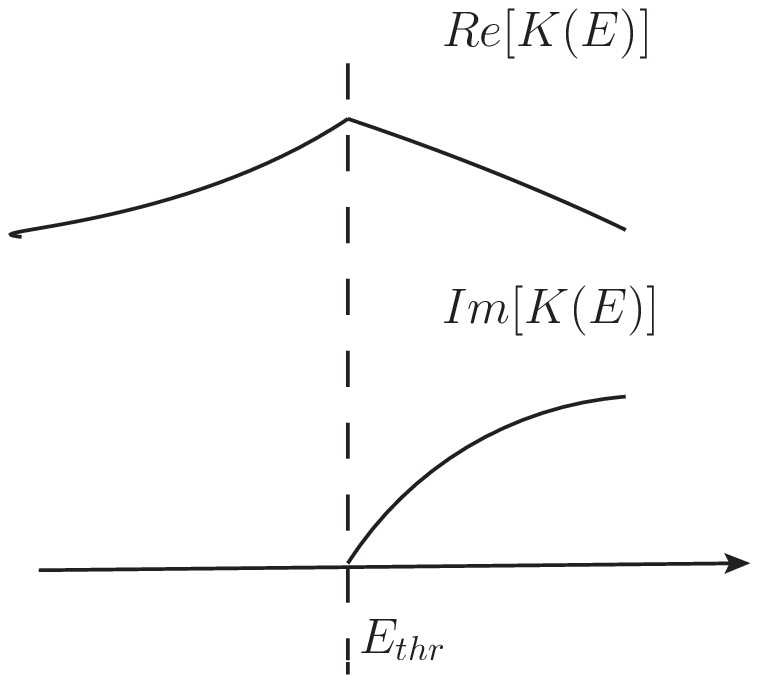}
\caption{a) Contours $C$ and $C'$ in the complex $E$ plane, exibiting analytic
properties of the integral (\ref{58}); b) Typical behaviour of the
$\textrm{Re}[K_n (E)]$ and $\textrm{Im}[ K_n(E)]$ given by Eq.
(\ref{59}).\label{fig:Contour}}
\end{figure}

At the end we study the analytic structure of the resonance denominator in
$\frac{1}{1-\hat K\, \hat \zeta}$, which in the one-channel case is given by
Eq. (\ref{53}). One can write $K_n (E)$ as the integral
\begin{equation}
K_n (E) = \frac{1}{4\pi^2}\int^\infty_{m_\pi}d\omega\,
\frac{k(\omega)\,\varphi^2 (k)}{M_n+\omega -E}\,, \label{58a}
\end{equation}
where $k(\omega) = \sqrt{\omega^2 - m^2_\pi}$. To display the analytic
properties of $K_n(E)$, we shall use the method, exploited in \cite{12},
Appendix E, i.e. we replace the integral in (\ref{58a}) as 1/2 of the contour
integral over the contour $C$, circumjacent to the  interval $[m_\pi,
 \infty]$, as shown in Fig.\ref{fig:Contour}, and take $E$ in the  physical region above
the contour $C$, $E\to E+ i\delta,\, \delta>0$. Deforming the contour, so that
it passes above and to the left of the point $E$ (contour $C'$ in
Fig.\ref{fig:Contour}), using Cauchy's theorem, one has the representation
\begin{equation}
K_n (E)= \frac{i}{4\pi}\, \varphi^2(k(E))\,k(E) + F_K(E)\equiv i\, a_K(E)\,
k(E) + b_K(E). \label{59}
\end{equation}
Here $k(E) = \sqrt{(E-M_n)^2- m^2_\pi}$, and $F_K (E)$ is analytic function in
the neighborhood of the threshold $E^{(n)}_{th} = M_n + m_\pi$. Hence
$\textrm{Re}[K_n (E)]$ acquires a negative contribution from the first term on
the r.h.s. of (\ref{59}) and behaves, as shown in Fig.\ref{fig:Contour}. This
behavior of $\textrm{Re}[K_n(E)]$ and $\textrm{Im}[K_n(E)]$ agrees with that,
obtained by numerical integration in Fig.\ref{fig:Zeta33 and K3}.

In a similar way one can write the form of $\zeta_{nm}(E)$
\begin{equation} \zeta_{nm} (E) = \int^\infty_0 \frac{p\,
dp^2}{4\pi^2 N_c} \frac{\chi^2(p)}{\left(\frac{p^2}{2\tilde M}-z\right) }
=\frac{2\tilde M}{4\,\pi N_c} \{ i\pi\,  p(z)\, \chi^2 (p(z)) + F_\zeta (z)\}
\equiv i\, a_\zeta(E)\, p(E) + b_\zeta(E)\label{60}\end{equation} where $p(z) =
\sqrt{2\tilde M z}, ~~ z=E-M_{n_2} - M_{n_3}, ~~\tilde M = \frac{M_{n_2}
M_{n_3}}{M_{n_2}+ M_{n_3}}$, and  $a_i, b_i\, (i=K,\zeta)$ are analytic and
positive functions near the thresholds.

From (\ref{59}), (\ref{60}) one can deduce, that the product $K_n(E)\,
\zeta_{nn}(E)$, is a real  analytic function in the complex plane of $E$ with
cuts, starting at thresholds $E_{th}^{(n)} = M_n + m_\pi$ and  $E_{n_2n_3} =
M_{n_2} + M_{n_3}$ and going to plus infinity. Thus the combination $(1
-K_n(E)\,\zeta_{nn}(E))$ is a real analytic function in the $E$ plane with
positive weights in the integrals (\ref{59}), (\ref{60}). Hence the  only
possibility for the zeros of $(1 -K_n(E)\,\zeta_{nn} (E))$ is on the real axis
below thresholds, or else on the next   sheets, which implies a standard
situation with possibility of  bound state or virtual state poles, or else
Breit-Wigner poles  in  $\frac{1}{1- K_n(E)\, \zeta_{nn}(E)}.$ From (\ref{59}),
(\ref{60}) one has
\begin{equation}
\frac{1}{1-K_n(E)\,\zeta_{nn}(E)} = \frac{1}{1-b_K\,b_\zeta + k\,p\, a_K\,
a_\zeta -i \,(k\,a_K\, b_\zeta + p\, a_\zeta\, b_K)}. \label{58}
\end{equation}
In the simple case, when the thresholds in $K_n(E)$ and $\zeta_{nn}(E)$
coincide the resonance factor acquires the form
\begin{equation}
\frac{1}{1-K_n(E)\,\zeta_{nn}(E)} = \frac{1}{A-i\,k\,B} \label{59a}
\end{equation}
with $B>0$. This form demonstrates the appearance of a virtual $(A>0)$ or  a
real $(A<0)$ pole.

\end{document}